\documentclass[preprint,12pt,fleqn]{elsarticle}
\usepackage{amssymb}
\usepackage{amsfonts}
\usepackage[english]{babel}
\usepackage{euscript}
\usepackage{bbm}
\usepackage{xfrac}

\journal{Journal of Geometry and Physics}

\newcommand{\fl}{\hspace*{-\mathindent}}
\newcommand{\textfrac}[2]{\textstyle{\frac{#1}{#2}}}
\newcommand{\eqref}[1]{(\ref{#1})}

\begin{document}

\begin{frontmatter}

\title{
Lax representations with non-removable parameters and
\\
integrable hierarchies  of PDEs via exotic cohomology of
\\
symmetry algebras
}

\author{Oleg I. Morozov}
\ead{morozov{\symbol{64}}agh.edu.pl}
\address{Faculty of Applied
  Mathematics, AGH University of Science and Technology,
  \\
  Al. Mickiewicza 30,
  Cracow 30-059, Poland,
\\
and
\\
Institute of Control Sciences of Russian Academy of Sciences,
  \\
  Profsoyuznaya 65, Moscow 117997, Russia
  }

\begin{abstract}
This paper develops the technique of constructing Lax representations for {\sc pde}s via non-central extensions
generated by non-triivial exotic 2-cocycles of their contact symmetry algebras. We show that the me\-thod is
applicable to the Lax re\-pre\-sen\-ta\-ti\-ons with non-removable spectral pa\-ra\-me\-ters. Also we demonstrate
that natural extensions of the symmetry algebras produce the integrable hierarchies associated to their
{\sc pde}s.
\end{abstract}


\end{frontmatter}




\section{Introduction}

Lax representations, also known as zero-curvature representations, Wahl\-qu\-ist--Es\-ta\-brook prolongation
structures, inverse scattering transformations, or differential co\-ve\-rings
\cite{KrasilshchikVinogradov1984,KrasilshchikVinogradov1989}, are a key feature of integrable partial
differential equations ({\sc pde}s). A number of important techniques for studying integrable {\sc pde}s
such as B\"acklund trans\-for\-ma\-ti\-ons, Dar\-boux transformations, recursion operators, nonlocal symmetries,
and nonlocal con\-ser\-va\-tion laws, are based on Lax representations. Lax representations with non-removable
(spectral) pa\-ra\-me\-ter are of special interest in the theory of integrable {\sc pde}s, see, e.g.,
\cite{AblowitzClarkson1991,Das1989,DoddFordy1983,TakhtadzhyanFaddeev1987}. The chal\-len\-ging unsolved
pro\-blem in this theory is to find conditions that are formulated in inherent terms of a {\sc pde} under study
and ensure existence of a Lax representation. Recently, an approach to this problem has been proposed in
\cite{Morozov2017,Morozov2018a}, where it was shown that for some {\sc pde}s their Lax representations can
be inferred from the second exotic cohomology group of the contact symmetry algebras of the {\sc pde}s.

The present paper provides an important supplement to the technique of \cite{Morozov2017,Morozov2018a}.
Na\-me\-ly, we show that Lax representations with non-removable pa\-ra\-me\-ters arise na\-tu\-ral\-ly from
non-central extensions of the symmetry algebras generated by nontrivial se\-cond exotic cohomology groups.
We consider here four equations: the hyper-CR equation for Ein\-stein--Weyl structures
 \cite{Kuzmina1967,Mikhalev1992,Pavlov2003,Dunajski2004}
\begin{equation}
u_{yy} = u_{tx} + u_y\,u_{xx} - u_x\,u_{xy},
\label{Pavlov_eq}
\end{equation}
the reduced quasi-classical self-dual Yang--Mills equation
\cite{FerapontovKhusnutdinova2004}
\begin{equation}
u_{yz} = u_{tx}+u_y\,u_{xx}-u_x\,u_{xy},
\label{FKh4}
\end{equation}
the four-dimensional  equation
\begin{equation}
u_{zz} = u_{tx} + u_z u_{xy} - u_x u_{yz},
\label{4D_UHE}
\end{equation}
 introduced in \cite{BogdanovPavlov2017},
and
the four-dimensional Mart{\'{\i}}nez Alonso--Shabat equation \cite{MartinezAlonsoShabat2004}
\begin{equation}
u_{ty} = u_z\,u_{xy} - u_y\,u_{xz}.
\label{MASh4}
\end{equation}
Equations \eqref{FKh4}  and \eqref{MASh4} are related by a B{\"a}cklund transformation,
\cite{KruglikovMorozov2015}, while their contact symmetry algebras are not isomorphic. Furthermore, they are
symmetry reductions of the quasi-classical self-dual Yang--Mills equation
\cite{ManakovSantini2006,ManakovSantini2014,BKMV2015,BogdanovPavlov2017}
\begin{equation}
u_{yz} = u_{tx}+u_x\,u_{zs}-u_z\,u_{xs}.
\label{MASh5}
\end{equation}

The 3-dimensional reduction of equation \eqref{4D_UHE} defined by substitution for $u_t=0$ produces the
universal hierarchy equation \cite{MartinezAlonsoShabat2002,MartinezAlonsoShabat2004}
\begin{equation}
u_{zz} = u_z u_{xy} - u_x u_{yz},
\label{UHE3}
\end{equation}
therefore we refer equation \eqref{4D_UHE} to as the {\it four-dimensional universal hierarchy equation}.

The Lax representations with non-removable parameters $\lambda$ for equations \eqref{Pavlov_eq} ---
\eqref{MASh4} are defined by systems
\begin{equation}
\left\{
\begin{array}{lcl}
v_t &=& (\lambda^2-\lambda\,u_x-u_y)\,v_x,
\\
v_y &=& (\lambda-u_x)\,v_x,
\end{array}
\right.
\label{Pavlov_eq_covering}
\end{equation}
\begin{equation}
\left\{
\begin{array}{lcl}
v_t &=& \lambda\,v_y-u_y\,v_x,
\\
v_z &=& (\lambda-u_x)\,v_x,
\end{array}
\right.
\label{FKh4_covering_lambda}
\end{equation}
\begin{equation}
\left\{
\begin{array}{lcl}
v_t&=& \lambda^2\,v_x -(\lambda\,u_x+u_z)\,v_y,
\\
v_z&=& \lambda\,v_x - u_x\,v_y,
\end{array}
\right.
\label{Pavlov_Stoilov_covering}
\end{equation}
and
\begin{equation}
\left\{
\begin{array}{lcl}
v_y &=& \lambda\,u_y\,v_x,
\\
v_z &=& \lambda\,(u_z\,v_x-v_t),
\end{array}
\right.
\label{MASh4_covering_lambda}.
\end{equation}
These systems were found in  \cite{Mikhalev1992,Pavlov2003,Dunajski2004},  \cite{FerapontovKhusnutdinova2004},
\cite{PavlovStoilov2017}, and   \cite{Morozov2014}, respectively.

The following structure distinguishes the contact symmetry algebras for equations
\eqref{Pavlov_eq} --- \eqref{MASh4}: these Lie algebras are semi-direct products
$\mathfrak{g}_\infty \rtimes \mathfrak{g}_\diamond$ of an infinite-di\-men\-si\-o\-nal ideal
$\mathfrak{g}_\infty$ and a non-Abelian  finite-dimensional Lie algebra $\mathfrak{g}_\diamond$. The
se\-cond exotic cohomology groups of the finite-dimensional subalgebras $\mathfrak{g}_\diamond$ turn out to be
non\-tri\-vi\-al for all the equations, and the corresponding nontrivial 2-cocycles produce non-central
ex\-ten\-si\-ons of their symmetry algebras $\mathfrak{g}_\infty \rtimes \mathfrak{g}_\diamond$.
We show that certain linear combinations of the Maurer--Cartan forms of the ex\-ten\-si\-ons define the Lax
re\-pre\-sen\-ta\-ti\-ons \eqref{Pavlov_eq_covering}, \eqref{FKh4_covering_lambda}, and
\eqref{MASh4_covering_lambda}, while the Lax representation \eqref{Pavlov_Stoilov_covering} can be revealed
via the same procedure applied twice.

The infinite-dimensional ideals $\mathfrak{g}_\infty$ of the symmetry algebras for the equations under
consideration admit series of natural extensions $\widehat{\mathfrak{g}}_\infty$ that preserve  the
actions of the finite-dimensional Lie algebras $\mathfrak{g}_\diamond$ as well as the nontrivial 2-cocycles from
the se\-cond exotic cohomology groups of $\mathfrak{g}_\diamond$. The nontrivial 2-cocycles generate non-central
extensions of the Lie algebras $\widehat{\mathfrak{g}}_\infty \rtimes \mathfrak{g}_\diamond$. The Maurer--Cartan
forms of the extensions provide Lax representations for integrable hierarchies associated with equations
\eqref{Pavlov_eq} --- \eqref{MASh4}. Thus we show that the integrable hierachies are invariantly and
intrinsically related to the equations under the study.


\section{Preliminaries}

All considerations in this paper are local. All functions are assumed to be real-analytic.


\subsection{Symmetries and differential coverings}

The presentation in this subsection closely follows
\cite{KrasilshchikVerbovetsky2011,KrasilshchikVerbovetskyVitolo2012},
see also \cite{KrasilshchikVinogradov1984,KrasilshchikVinogradov1989,VK1999}.
Let $\pi \colon \mathbb{R}^n \times \mathbb{R}^m \rightarrow \mathbb{R}^n$,
$\pi \colon (x^1, \dots, x^n, u^1, \dots, u^m) \mapsto (x^1, \dots, x^n)$, be a trivial bundle, and
$J^\infty(\pi)$ be the bundle of its jets of the infinite order. The local coordinates on $J^\infty(\pi)$ are
$(x^i,u^\alpha,u^\alpha_I)$, where $I=(i_1, \dots, i_n)$ are multi-indices, and for every local section
$f \colon \mathbb{R}^n \rightarrow \mathbb{R}^n \times \mathbb{R}^m$ of $\pi$ the corresponding infinite jet
$j_\infty(f)$ is a section $j_\infty(f) \colon \mathbb{R}^n \rightarrow J^\infty(\pi)$ such that
$u^\alpha_I(j_\infty(f))
=\displaystyle{\frac{\partial ^{\#I} f^\alpha}{\partial x^I}}
=\displaystyle{\frac{\partial ^{i_1+\dots+i_n} f^\alpha}{(\partial x^1)^{i_1}\dots (\partial x^n)^{i_n}}}$.
We put $u^\alpha = u^\alpha_{(0,\dots,0)}$. Also, we will simplify notation in the following way, e.g., in the
case of $n=4$, $m=1$: we denote $x^1 = t$, $x^2= x$, $x^3= y$, $x^4=z$ and
$u^1_{(i,j,k,l)}=u_{{t \dots t}{x \dots x}{y \dots y}{z \dots z}}$ with $i$  times $t$,
$j$  times $x$, $k$  times $y$, and $l$ times $z$.

The  vector fields
\[
D_{x^k} = \frac{\partial}{\partial x^k} + \sum \limits_{\# I \ge 0} \sum \limits_{\alpha = 1}^m
u^\alpha_{I+1_{k}}\,\frac{\partial}{\partial u^\alpha_I},
\qquad k \in \{1,\dots,n\},
\]
$(i_1,\dots, i_k,\dots, i_n)+1_k = (i_1,\dots, i_k+1,\dots, i_n)$,  are called {\it total derivatives}.
They com\-mu\-te everywhere on
$J^\infty(\pi)$:  $[D_{x^i}, D_{x^j}] = 0$.

The {\it evolutionary vector field} associated to an arbitrary vector-valued smooth function
$\varphi \colon J^\infty(\pi) \rightarrow \mathbb{R}^m $ is the vector field
\[
\mathbf{E}_{\varphi} = \sum \limits_{\# I \ge 0} \sum \limits_{\alpha = 1}^m
D_I(\varphi^\alpha)\,\frac{\partial}{\partial u^\alpha_I}
\]
with $D_I=D_{(i_1,\dots\,i_n)} =D^{i_1}_{x^1} \circ \dots \circ D^{i_n}_{x^n}$.

A system of {\sc pde}s $F_r(x^i,u^\alpha_I) = 0$ of the order $s \ge 1$ with $\# I \le s$,
$r \in \{1,\dots, R\}$ for some $R \ge 1$,
defines the submanifold
$\EuScript{E}=\{(x^i,u^\alpha_I)\in J^\infty(\pi)\,\,\vert\,\,D_K(F_r(x^i,u^\alpha_I))=0,\,\,\# K\ge 0\}$
in $J^\infty(\pi)$.

A function $\varphi \colon J^\infty(\pi) \rightarrow \mathbb{R}^m$ is called a {\it (generator of an
infinitesimal) symmetry} of equation $\EuScript{E}$ when $\mathbf{E}_{\varphi}(F) = 0$ on $\EuScript{E}$. The
symmetry $\varphi$ is a solution to the {\it defining system}
\begin{equation}
\ell_{\EuScript{E}}(\varphi) = 0,
\label{defining_eqns}
\end{equation}
where $\ell_{\EuScript{E}} = \ell_F \vert_{\EuScript{E}}$ with the matrix differential operator
\[
\ell_F = \left(\sum \limits_{\# I \ge 0}\frac{\partial F_r}{\partial u^\alpha_I}\,D_I\right).
\]
The {\it symmetry algebra} $\mathrm{Sym} (\EuScript{E})$ of equation $\EuScript{E}$ is the linear space of
solutions to  (\ref{defining_eqns}) endowed with the structure of a Lie algebra over $\mathbb{R}$ by the
{\it Jacobi bracket} $\{\varphi,\psi\} = \mathbf{E}_{\varphi}(\psi) - \mathbf{E}_{\psi}(\varphi)$.
The {\it algebra of contact symmetries} $\mathrm{Sym}_0 (\EuScript{E})$ is the Lie subalgebra of $\mathrm{Sym} (\EuScript{E})$
defined as $\mathrm{Sym} (\EuScript{E}) \cap J^1(\pi)$.

Consider $\EuScript{W} = \mathbb{R}^\infty$ with  coordinates $w^s$, $s \in  \mathbb{N}_{0}$. Locally,
an (infinite-di\-men\-si\-o\-nal)  {\it differential covering} of $\EuScript{E}$ is a trivial bundle
$\tau \colon J^\infty(\pi) \times \EuScript{W} \rightarrow J^\infty(\pi)$
equipped with {\it extended total derivatives}
\[
\widetilde{D}_{x^k} = D_{x^k} + \sum \limits_{ s =0}^\infty
T^s_k(x^i,u^\alpha_I,w^j)\,\frac{\partial }{\partial w^s}
\]
such that $[\widetilde{D}_{x^i}, \widetilde{D}_{x^j}]=0$ for all $i \not = j$ whenever
$(x^i,u^\alpha_I) \in \EuScript{E}$. Define
the partial derivatives of $w^s$ by  $w^s_{x^k} =  \widetilde{D}_{x^k}(w^s)$.  This yields the system of
{\it covering equations}
\begin{equation}
w^s_{x^k} = T^s_k(x^i,u^\alpha_I,w^j)
\label{WE_prolongation_eqns}
\end{equation}
that is compatible whenever $(x^i,u^\alpha_I) \in \EuScript{E}$.

Dually, the covering is defined by the {\it Wahlquist--Estabrook forms}
\begin{equation}
d w^s - \sum \limits_{k=1}^{m} T^s_k(x^i,u^\alpha_I,w^j)\,dx^k
\label{WEfs}
\end{equation}
as follows: when $w^s$  and $u^\alpha$ are considered to be functions of $x^1$, ... , $x^n$, forms \eqref{WEfs}
are equal to zero whenever system \eqref{WE_prolongation_eqns} holds.


\subsection{Exotic cohomology of Lie algebras}

For a Lie algebra
 $\mathfrak{g}$ over $\mathbb{R}$, its representation $\rho \colon \mathfrak{g} \rightarrow \mathrm{End}(V)$,
and $k \ge 1$
let $C^k(\mathfrak{g}, V) =\mathrm{Hom}(\Lambda^k(\mathfrak{g}), V)$
be the space of all $k$--linear skew-symmetric mappings from $\mathfrak{g}$ to $V$. Then
the Chevalley--Eilenberg differential
complex
\[
V=C^0(\mathfrak{g}, V) \stackrel{d}{\longrightarrow} C^1(\mathfrak{g}, V)
\stackrel{d}{\longrightarrow} \dots \stackrel{d}{\longrightarrow}
C^k(\mathfrak{g}, V) \stackrel{d}{\longrightarrow} C^{k+1}(\mathfrak{g}, V)
\stackrel{d}{\longrightarrow} \dots
\]
is generated by the differential $d \colon \theta \mapsto d\theta$ such that
\[
d \theta (X_1, ... , X_{k+1}) =
\sum\limits_{q=1}^{k+1}
(-1)^{q+1} \rho (X_q)\,(\theta (X_1, ... ,\hat{X}_q, ... ,  X_{k+1}))
\]
\[
\quad
+\sum\limits_{1\le p < q \le k+1} (-1)^{p+q+1}
\theta ([X_p,X_q],X_1, ... ,\hat{X}_p, ... ,\hat{X}_q, ... ,  X_{k+1}).
\]
The cohomology groups of the complex $(C^{*}(\mathfrak{g}, V), d)$ are referred to as
the {\it cohomology groups of the Lie algebra} $\mathfrak{g}$ {\it with coefficients in the representation}
$\rho$. For the trivial representation $\rho_0 \colon \mathfrak{g} \rightarrow \mathbb{R}$,
$\rho_0 \colon X \mapsto 0$, the cohomology groups are denoted by
$H^{*}(\mathfrak{g})$.

Consider a Lie algebra $\mathfrak{g}$ over $\mathbb{R}$ with non-trivial first cohomology group
$H^1(\mathfrak{g})$ and take a closed 1-form $\alpha$ on $\mathfrak{g}$ such that $[\alpha] \neq 0$.
Then for any $c \in \mathbb{R}$
define new differential
$d_{c \alpha} \colon C^k(\mathfrak{g},\mathbb{R}) \rightarrow C^{k+1}(\mathfrak{g},\mathbb{R})$ by
the formula
\[
d_{c \alpha} \theta = d \theta - c \,\alpha \wedge \theta.
\]
From  $d\alpha = 0$ it follows that
$d_{c \alpha} ^2=0$. The cohomology groups of the complex
\[
C^1(\mathfrak{g}, \mathbb{R})
\stackrel{d_{c \alpha}}{\longrightarrow}
\dots
\stackrel{d_{c \alpha}}{\longrightarrow}
C^k(\mathfrak{g}, \mathbb{R})
\stackrel{d_{c \alpha}}{\longrightarrow}
C^{k+1}(\mathfrak{g}, \mathbb{R})
\stackrel{d_{c \alpha}}{\longrightarrow} \dots
\]
are referred to as the {\it exotic} {\it cohomology groups}  \cite{Novikov2002,Novikov2005} of $\mathfrak{g}$
and denoted by $H^{*}_{c\alpha}(\mathfrak{g})$.


\section{Hyper-CR equation}


\subsection{Contact symmetries}

Denote by $\EuScript{E}_1$ the hyper-CR equation \eqref{Pavlov_eq}.
Direct computations\footnote[1]{We carried out computations of generators of contact symmetries and their
commutator tables in the {\it Jets} software \cite{Jets}.} show that the Lie algebra
$\mathrm{Sym}_0(\EuScript{E}_1)$ is generated by functions
\[
W_0(A) = -A\,u_t -\left(x\,A^{\prime} +\textfrac{1}{2}\,y^2\,A^{\prime\prime}\right)\,u_x
-y\,A^{\prime}\, u_y+u\,A^{\prime} +x\,y\,A^{\prime\prime} + \textfrac{1}{6}\,y^3\,A^{\prime\prime\prime},
\]
\[
W_1(A) = - y\,A^{\prime}\,u_x- A\,u_y+x\,A^{\prime} + \textfrac{1}{2}\,y^2\,A^{\prime\prime},
\]
\[
W_2(A) = -A\,u_x+y\,A^{\prime},
\]
\[
W_3(A) = A,
\]
\[
X_0 = -2\,x\,u_x-y\,u_y+3\,u,
\]
\[
X_1 = - y\,u_x+2\,x
\]
where $A=A(t)$ and $B=B(t)$ below are arbitrary functions of $t$. The commutators of the generators are given
by equations
\begin{equation}
\left\{
\begin{array}{lcl}
\{W_i(A), W_j(B)\}  &=&  W_{i+j}(A\,B^{\prime} - B\,A^{\prime}),
\\
\{X_i, W_k(A)\} &=& -k\,W_{k+i}(A),
\\
\{X_0, X_1\} &=& -X_1,
\end{array}
\right.
\label{Pavlov_eq_commutator_table}
\end{equation}
where $W_k(A)=0$ for $k > 3$. From equations \eqref{Pavlov_eq_commutator_table} it follows that the contact
sym\-met\-ry algebra of $\EuScript{E}_1$ is the semi-direct product
$\mathrm{Sym}_0(\EuScript{E}_1) =\mathfrak{p}_{4,\infty} \rtimes \mathfrak{p}_\diamond$
of the two-di\-men\-si\-o\-nal non-Abelian Lie algebra $\mathfrak{p}_\diamond =\langle X_0, X_1 \rangle$
and the infinite-dimensional ideal
$\mathfrak{p}_{4,\infty} = \langle W_k(A) \,\,\,\vert\,\,\, 0\le k \le 3 \rangle$. The ideal, in its turn, is
isomorphic to the tensor product $\mathbb{R}_3[h_0] \otimes \mathfrak{w}[t]$ of the four-dimensional
commutative associative algebra of truncated polynomials $\mathbb{R}_3[h_0] = \mathbb{R}[h_0]/(h_0^4 = 0)$ and
the Lie algebra $\mathfrak{w}[t] = \langle \, t^i \partial_t \,\,\,\vert\,\,\, i \in \mathbb{N}_{0} \,\rangle$.

Consider the basis of $\mathrm{Sym}_0(\EuScript{E}_1)$ given by generators $X_0$, $X_1$, and
$\displaystyle{\frac{1}{i!}} W_k(t^i)$ with $i \in \mathbb{N}_{0}$, $k \in \{0, \dots,  3\}$. Define the
Maurer--Cartan forms   $\alpha_0$, $\alpha_1$, $\theta_{k,i}$  for $\mathrm{Sym}_0(\EuScript{E}_1)$ as the dual
forms to this basis: $\alpha_m(X_j) = \delta_{ij}$, $\alpha_m(W_k(t^i)) = 0$, $\theta_{k,i}(X_j) = 0$,
$\theta_{k,i}(W_{k^\prime}(t^{i^\prime})) = i!\,\delta_{kk^{\prime}} \delta_{ii^\prime}$.
Put
\begin{equation}
\Theta = \sum \limits_{k=0}^{3} \sum \limits_{m=0}^{\infty} \frac{h_0^kh_1^m}{m!}\,\theta_{k,m},
\label{Theta_Pavlov}
\end{equation}
where $h_0$ and $h_1$ are formal parameters such that $dh_i=0$ and $h_0^k =0$ when $k>3$. Denote by $\nabla_0$
the derivative with respect to $h_0$ in $\mathbb{R}_3[h_0]$ and put $\nabla_1 =\partial_{h_1}$. Then the
commutator table \eqref{Pavlov_eq_commutator_table} gives the {\it Maurer--Cartan structure equations}
\begin{equation}
\left\{
\begin{array}{lcl}
d\alpha_0 &=& 0,
\\
d\alpha_1 &=& \alpha_0 \wedge \alpha_1,
\\
d \Theta &=& \nabla_1 (\Theta) \wedge \Theta + (h_0\,\alpha_0 + h_0^2\,\alpha_1) \wedge \nabla_0 (\Theta)
\end{array}
\right.
\label{structure_equations_for_Pavlov_eq}
\end{equation}
of $\mathrm{Sym}_0(\EuScript{E}_1)$.

\vskip 7 pt
\noindent
{\sc Remark}.
The last equation in system \eqref{structure_equations_for_Pavlov_eq} is actually a short form of an infinite
 system that includes four series of equations for $\theta_{0,i}$, ... , $\theta_{3,i}$. For example, the first
equations from each series are given as follows:
\begin{equation}
\fl
\left\{
\begin{array}{lcl}
d\theta_{0,0} &=& \theta_{0,1}\wedge \theta_{0,0},
\\
d\theta_{1,0} &=& (\alpha_0+\theta_{0,1}) \wedge \theta_{1,0} + \theta_{1,1} \wedge \theta_{0,0},
\\
d\theta_{2,0} &=& (2\,\alpha_0+\theta_{0,1}) \wedge \theta_{2,0}
+ (\alpha_{1}+ \theta_{1,1}) \wedge \theta_{1,0}
+ \theta_{2,1} \wedge \theta_{0,0},
\\
d\theta_{3,0} &=& (3\,\alpha_0+\theta_{0,1}) \wedge \theta_{3,0}
+ (2\,\alpha_{1} + \theta_{1,1})\wedge \theta_{2,0}
+ \theta_{2,1} \wedge \theta_{1,0}
\\
&&
+ \theta_{3,1} \wedge \theta_{0,0}.
\end{array}
\right.
\label{first_structure_equations_for_Pavlov_eq}
\end{equation}
Equations for all the other forms $\theta_{k,i}$ can be obtained from equations
\eqref{first_structure_equations_for_Pavlov_eq} by the pro\-ce\-du\-re of normal prolongation, \cite{Cartan1,Cartan4}.
For the purposes of the present paper we need ex\-pli\-cit expressions for the first forms $\theta_{k,0}$ of each
series only. To shorten the notation we will write $\theta_k$ instead of $\theta_{k,0}$ in this section
or instead of $\theta_{k,0,0}$ in the next sections.
\hfill $\diamond$


\subsection{Second exotic cohomology group and non-central extension}

From the structure equations \eqref{structure_equations_for_Pavlov_eq}
we have
\vskip 7 pt
\noindent
{\sc Proposition 1}.
{\it
$H^1(\mathrm{Sym}_0(\EuScript{E}_1)) = \langle [\alpha_0]\rangle=\langle \alpha_0\rangle$
and
\[
H^2_{c\,\alpha_0}(\mathfrak{p}_\diamond)  =
\left\{
\begin{array}{lcl}
\langle [\alpha_0 \wedge \alpha_1]\rangle, &~~~& c =1,
\\
\{[0]\}, && c \neq 1.
\end{array}
\right.
\]
Furthermore,
$H^2_{\alpha_0}(\mathfrak{p}_\diamond) \subseteq H^2_{\alpha_0}(\mathrm{Sym}_0(\EuScript{E}_1))$. Hence the
nontrivial 2-cocycle $\alpha_0 \wedge \alpha_1$ of the differential $d_{\alpha_0}$ defines a non-central
extension $\widehat{\mathfrak{p}}_\diamond$ of the Lie algebra $\mathfrak{p}_\diamond$ and thus a non-central
extension $\mathfrak{p}_{4,\infty} \rtimes \widehat{\mathfrak{p}}_\diamond$ of the Lie algbera
$\mathrm{Sym}_0(\EuScript{E}_1)$. The additional Maurer--Cartan form $\sigma$ for the extended Lie algebra is
a solution to $d_{\alpha_0} \sigma = \alpha_0 \wedge \alpha_1$, that is, to equation
\begin{equation}
d \sigma = \alpha_0 \wedge \sigma + \alpha_0 \wedge \alpha_1.
\label{extension_se_1}
\end{equation}
This equation is  compatible with the structure equations \eqref{structure_equations_for_Pavlov_eq}
of the Lie algebra $\mathrm{Sym}_0(\EuScript{E}_1)$.
}
\hfill $\Box$


\subsection{Maurer--Cartan forms and Lax representation}
\label{Pavlov_MCfs_subsection}

We can compute the Maurer--Cartan forms $\alpha_i$, $\theta_{k}$, and $\sigma$ via two approaches. The first
one is to integrate equations \eqref{structure_equations_for_Pavlov_eq}, \eqref{extension_se_1} step by step.
Each integration gives certain num\-ber of  new coordinates (the `integration constants') to express the new
form, while it is not clear how these coordinates are related to the coordinates of $\EuScript{E}_1$.
For example, from the first two equations of system \eqref{structure_equations_for_Pavlov_eq} and equation
\eqref{extension_se_1} we obtain
\[
\alpha_0 = dq,
\quad
\alpha_1 = - \mathrm{e}^q\, ds,
\quad
\sigma  = \mathrm{e}^q\,(dv-q \,ds),
\]
where $q$, $s$, and $v$ are free parameters\footnote[2]{we put $\alpha_1=-\mathrm{e}^q\,ds$ instead of the natural choice
$\alpha_1=\mathrm{e}^q\,ds$  to simplify the computations below.}. The second approach to computing  the
Maurer--Cartan forms is to use Cartan's method of equivalence,
\cite{Cartan1,Cartan2,Cartan3,Cartan4,Olver1995,FelsOlver1998},
see details and examples of applying the method to symmetries of {\sc pde}s in \cite{Morozov2002,Morozov2006}.
For the sym\-met\-ry algebra of equation \eqref{Pavlov_eq} the combination of both techniques shows that
\begin{enumerate}
\item[{\it(i)}] $\theta_{0}$ is a multiple of $dt$, $\theta_{1}$
belongs to the algebraic ideal of 1-forms generated by $dy$, $dt$,
$\theta_{2}$ belongs to the ideal  generated by  $dx$, $dy$, $dt$;
\item[{\it(ii)}] $\theta_{3}$ is a multiple
of the contact form $du -u_t\,dt - u_x\,dx - u_y\,dy$.
\end{enumerate}
Using {\it(i)} we have
$\theta_{0} = r\,dt$,
$\theta_{1} = \mathrm{e}^q\,r\,(dy +p_1\,dt)$,
$\theta_{2} = \mathrm{e}^{2q}\,r\,(dx + (p_1-s)\,dy + p_2\,dt)$,
with new parameters $r \neq 0$, $p_1$,  $p_2$, while
{\it (ii)} then gives $p_1 = - u_x+2\,s$, $p_2  =- u_y -s\,u_x+s^2$, and
\begin{equation}
\theta_{3} = \mathrm{e}^{3q}\,r\,(du -u_t\,dt - u_x\,dx - u_y\,dy).
\label{Pavlov_theta_3_0}
\end{equation}
Consider the linear combination
\[
\fl
\qquad
\sigma - \theta_{2} =\mathrm{e}^q\,\left(dv - q \,ds -\mathrm{e}^q\,r\,\left(dx
+(s^2-s\,u_x-u_y)\,dt+(s-u_x)\,dy\right)\right)
\]
and assume that $u$ and $v$ are functions of $t$, $x$, $y$. Then $\sigma - \theta_{2}=0$ implies
$q = v_s$,
$r = v_x\,\exp (-v_s)$.
After this change of notation we obtain the Wahlquist--Estabrook form
\[
\fl
\qquad
\sigma - \theta_{2} =\mathrm{e}^{v_s}\,\left(dv -v_s\,ds -v_x\,(dx +(s^2-s\,u_x-u_y)\,dt
+(s-u_x)\,dy
)\right)
\]
of the Lax representation
\[
\left\{
\begin{array}{lcl}
v_t &=& (s^2-s\,u_x-u_y)\,v_x,
\\
v_y &=& (s-u_x)\,v_x.
\end{array}
\right.
\]
This system differs from \eqref{Pavlov_eq_covering} by notation.


\subsection{Integrable hierarchy associated to hyper-CR equation}
\label{Dunajski_hierarchy_subsection}

The Lie algebra $\mathfrak{p}_{4} = \mathfrak{p}_{4,\infty} \rtimes \mathfrak{p}_\diamond$ admits a sequence of
natural extensions $\mathfrak{p}_{n+1} = \mathfrak{p}_{n+1,\infty} \rtimes \mathfrak{p}_\diamond$, $n\ge 4$,
where $\mathfrak{p}_{n+1,\infty} =  \mathbb{R}_{n}[h_0] \otimes \mathfrak{w}[t]$ and
$\mathbb{R}_{n}[h_0]=\mathbb{R}[h_0]/(h_0^{n+1}=0)$.
These extensions are defined by the structure equations of the same form
\eqref{structure_equations_for_Pavlov_eq},
where now  we put
\begin{equation}
\Theta = \sum \limits_{k=0}^{n} \sum \limits_{m=0}^{\infty} \frac{h_0^k\,h_1^m}{m!}\,\theta_{k,m},
\label{Theta_Pavlov_n}
\end{equation}
instead of \eqref{Theta_Pavlov} and assume $h_0^k =0$ for $k > n$.  The finite-dimensional part
$\mathfrak{p}_\diamond$ in all the algebras $\mathfrak{p}_{n+1}$ is the same, and for each $n \ge 4$
we have $H^2_{\alpha_0}(\mathfrak{p}_\diamond) \subseteq H^2_{\alpha_0}(\mathfrak{p}_{n+1})$. Hence the
nontrivial 2-cocycle $\alpha_0 \wedge \alpha_1$ of the differential $d_{\alpha_0}$ defines a non-central
extension $\widehat{\mathfrak{p}}_{n+1} = \mathfrak{p}_{n+1,\infty} \rtimes \widehat{\mathfrak{p}}_\diamond$ of
the Lie algebra $\mathfrak{p}_{n+1}$. The structure equations for $\widehat{\mathfrak{p}}_{n+1}$ are given by
\eqref{structure_equations_for_Pavlov_eq}, \eqref{Theta_Pavlov_n}, and \eqref{extension_se_1}.

For a fixed $n \ge 4$ we can find forms $\theta_{k}$ with $0 \le k \le n$ by integration of the structure
equations of $\mathfrak{p}_{n+1}$. In the next subsections we give examples of such computations. There we
alter notation as follows: $t \mapsto t_0$,  $y \mapsto t_1$, $x \mapsto t_2$.


\subsubsection{Case $n=4$.}\label{Pavlov_eq_hierarchy_subsection_n=5}

While the 1-forms $\alpha_0$, $\alpha_1$,  $\sigma$, $\theta_{0}$, and  $\theta_{1}$ are the same
as in section \ref{Pavlov_MCfs_subsection}, instead of \eqref{Pavlov_theta_3_0} we have now
\[
\theta_{3} = \mathrm{e}^{3q}\,r\,
\left(
dt_3 + (p_1-2\,s)\,dt_2 + (p_2-s\,p_1+s^2)\,dt_1 + p_3 \,dt_0
\right).
\]
Next integration gives
\[
\theta_{4} = \mathrm{e}^{4q}\,r\,
(dt_4 + (p_1-3\,s)\,dt_3 + (p_2-2\,s\,p_1+3\,s^2)\,dt_2
\]
\[
\qquad\qquad
+(p_3-s\,p_2+s^2\,p_1-s^3)\,dt_1+ p_4 \,dt_0).
\]
We enforce $\theta_{4}$ to be the contact form
$q^4\,r\,(du-\sum \limits_{i=0}^{3} u_{t_i}\,dt_i)$,
that is, we put
$t_4=u$,
$p_1 = -u_{t_3} +3\,s$,
$p_2 = -u_{t_2} -2\,s\,u_{t_3} +3\,s^2$,
$p_3 = -u_{t_1} -s\,u_{t_2} + s^2\,u_{t_3}+s^3$,
$p_4=-u_{t_0}$.
Then we consider the linear combination
\[
\fl
\sigma -\theta_{3} = \mathrm{e}^q\,(dv - q \,ds -\mathrm{e}^{2q}\,r\,(dt_3+(s-u_{t_3})\,dt_2
+(s^2-s\,u_{t_3} -u_{t_2})\,dt_1
\]
\[
\fl\qquad\qquad
+(s^3-s^2\,u_{t_3}-s^2\,u_{t_2}-u_{t_1})\,dt_0).
\]
Substituting  for  $q=v_s$, $r=v_{t_3}\,\mathrm{e}^{-v_s}$ yields
\[
\fl
\sigma -\theta_{0} = \mathrm{e}^{v_s}\,(dv - v_s \,ds -v_{t_3}\,(dt_3+(s-u_{t_3})\,dt_2
+(s^2-s\,u_{t_3} -u_{t_2})\,dt_1
\]
\begin{equation}
\fl\qquad\qquad
+(s^3-s^2\,u_{t_3}-s^2\,u_{t_2}-u_{t_1})\,dt_0).
\label{WE_form_of_Pavlov_eq_hierarchy_5}
\end{equation}
We consider $u$ as a function of $t_0$, ..., $t_3$ and $v$ as a function of $t_0$, ..., $t_3$, $s$.
Then equation \eqref{WE_form_of_Pavlov_eq_hierarchy_5} defines the Wahlquist--Estabrook
form for the Lax re\-pre\-sen\-ta\-ti\-on
\[
\left\{
\begin{array}{lcl}
v_{t_2}&=&(s-u_{t_3})\,v_{t_3},
\\
v_{t_1}&=&(s^2-s\,u_{t_3} -u_{t_2})\,v_{t_3},
\\
v_{t_0}&=&(s^3-s^2\,u_{t_3}-s^2\,u_{t_2}-u_{t_1})\,v_{t_3}
\end{array}
\right.
\]
of a system of {\sc pde}s.

To make the structures of this Lax representation and of the defined system more tractable we alter
notation by substituting  $t_i \mapsto x_{3-i}$ for $i \in \{0, \dots, 3\}$.  The obtained system
\[
\left\{
\begin{array}{lcl}
v_{x_1}&=&(s-u_{x_0})\,v_{x_0},
\\
v_{x_2}&=&(s^2-s\,u_{x_0} -u_{x_1})\,v_{x_0},
\\
v_{x_3}&=&(s^3-s^2\,u_{x_0}-s^2\,u_{x_1}-u_{x_2})\,v_{x_0}
\end{array}
\right.
\]
is compatible whenever there holds
\begin{equation}
u_{x_1x_1} = u_{x_0x_2}+u_{x_1}\,u_{x_0x_0}-u_{x_0}\,u_{x_0x_1},
\label{Pavlov_eq_renamed}
\end{equation}
\begin{equation}
u_{x_1x_2} = u_{x_0x_3}+u_{x_2}\,u_{x_0x_0}-u_{x_0}\,u_{x_0x_2},
\label{FKh_eq_renamed}
\end{equation}
\begin{equation}
u_{x_1x_3} = u_{x_2x_2} +u_{x_1}\,u_{x_0x_2}-u_{x_2}\,u_{x_0x_1}.
\label{UHE4_renamed}
\end{equation}
Equations \eqref{Pavlov_eq_renamed}, \eqref{FKh_eq_renamed}, \eqref{UHE4_renamed} differ from equations
\eqref{Pavlov_eq}, \eqref{FKh4}, \eqref{4D_UHE}, respectively, by notation.


\subsubsection{Case $n>4$.}

For the Lie algebra $\widehat{\mathfrak{p}}_{n+1}$ with fixed $n>4$ the results of com\-pu\-ta\-ti\-ons are the
following. Put $p_0=1$ and for $i \ge 0$, $j \in \{0, \dots, i\}$ define polynomials
$P_{ij}=P_{ij}(s)$
of variable $s$ by the formula
\begin{equation}
P_{ij} = \sum \limits_{k=0}^{j} (-1)^k\,
\left(
\begin{array}{c}
i-j+k-1\\k
\end{array}
\right)
\,p_{j-k}\,s^k.
\label{P_definition}
\end{equation}
Coefficients of $P_{ij}$ depend on parameters $p_1$, ..., $p_j$.
Then we have expressions
\[
\theta_{k} = \mathrm{e}^{kq}\,r\,\sum \limits_{j=0}^{k} P_{kj}\,dt_{k-j}
\]
for forms $\theta_{k}$ with $k \in \{0, \dots, n\}$.
We put $t_n=u$, solve the triangular linear system of equations
\[
P_{n,n-i} = - u_{t_{i}}, \qquad i \in \{0, \dots, n-1\}
\]
with respect to unknowns $p_1$, $p_2$, ... , $p_n$, and then alter notation by substituting
$t_i = x_{n-i}$ for $i \in \{0, \dots, n-1\}$.
This yields
\[
\theta_{n} = \mathrm{e}^{n q} r\,\left(du - \sum \limits_{i=0}^{n-1} u_{x_i} dx_i\right).
\]
Then we consider the linear combination $\sigma -\theta_{n-1}$ and put $q = v_s$,
$r = v_{x_0}\,\mathrm{e}^{(1-n)\,v_s}$. This produces the Wahlquist--Estabrook form
\[
\fl
\qquad
\sigma -\theta_{n-1} = \mathrm{e}^{v_s}\,
\left(dv -v_s\,ds
- v_{x_0}\,dx_0
- \sum \limits_{i=1}^{n-1} \left(s^i -\sum \limits_{j=0}^{i-1}\,s^{i-j-1}\,u_{x_j}\right)\,v_{x_0}\, dx_i
\right)
\]
for the Lax representation
\begin{equation}
\left\{
\begin{array}{lcl}
v_{x_1} &=& (s-u_{x_0})\,v_{x_0},
\\
v_{x_2} &=& (s^2-s\,u_{x_0}-u_{x_1})\,v_{x_0},
\\
&& \dots
\\
v_{x_i} &=&\left(s^i -\sum \limits_{j=0}^{i-1}\,s^{i-j-1}\,u_{x_j}\right)\,v_{x_0},
\\
&& \dots
\\
v_{x_{n-1}} &=&\left(s^{n-1} -s^{n-2} \,u_{x_0} -s^{n-3} \,u_{x_1} - \dots- s\,u_{x_{n-3}} -
u_{x_{n-2}}
\right)\,v_{x_0}.
\end{array}
\right.
\label{Pavlov_eq_hierarchy_general_covering}
\end{equation}
Denote by $\EuScript{H}_{n-1}$ the compatibility conditions for system
\eqref{Pavlov_eq_hierarchy_general_covering}. Then $\EuScript{H}_2$ is given by the single equation
\eqref{Pavlov_eq_renamed}, this equation supplemented by equations \eqref{FKh_eq_renamed},
\eqref{UHE4_renamed} defines  $\EuScript{H}_3$,  system $\EuScript{H}_4$
consists of equations from  $\EuScript{H}_3$  supplemented by  equations
\[
u_{x_0x_4} = u_{x_2x_2}+u_{x_0}\,u_{x_0x_3}-u_{x_3}\,u_{x_0x_0}
+u_{x_1}\,u_{x_0x_2}-u_{x_2}\,u_{x_0x_1},
\]
\[
u_{x_1x_4} =u_{x_2x_3}+u_{x_0}\,u_{x_0x_3}-u_{x_3}\,u_{x_0x_1},
\]
\[
u_{x_2x_4} =u_{x_3x_3} + u_{x_2}\,u_{x_0x_3}-u_{x_3}\,u_{x_0x_2},
\]
etc., system $\EuScript{H}_{n-1}$ consists of equations from  $\EuScript{H}_{n-2}$  supplemented by  equations
\[
u_{x_{i-1}x_n}=u_{x_ix_{n-1}}+u_{x_0}u_{x_0x_{n-1}}-u_{x_{n-1}}u_{x_0x_{i-1}},
\qquad i \in \{1, \dots, n-2\},
\]
where $u_{x_ix_{n-1}}$ are replaced by the right-hand sides of the equations from $\EuScript{H}_{n-2}$.
The  Lax representations \eqref{Pavlov_eq_hierarchy_general_covering} and systems $\EuScript{H}_{n}$ were
introduced in \cite{Dunajski2004}, see also \cite{Pavlov2003,BogdanovPavlov2017}.


\section{Reduced quasi-classical self-dual Yang--Mills equation}

Equation \eqref{FKh4} differs by notation from equation \eqref{FKh4_renamed} in the hierarchies $\EuScript{H}_m$
with $m \ge 3$. While the symmetry algebra $\mathrm{Sym}_0(\EuScript{E}_2)$ of equation (\ref{FKh4}) has more
complicated structure than $\mathrm{Sym}_0(\EuScript{E}_1)$, we show that the Lax representation
\eqref{FKh4_covering_lambda} as well as the integrable hierarchy associated to \eqref{FKh4} can be inferred from
the Maurer--Cartan forms of the non-central extension ge\-ne\-ra\-ted by the nontrivial exotic 2-cocycle of
$\mathrm{Sym}_0(\EuScript{E}_2)$.


\subsection{Contact symmetries}

The Lie algebra $\mathrm{Sym}_0(\EuScript{E}_2)$ admits generators
\begin{eqnarray*}
W_0(A) &=& -(A_z\,x+A_t\,y)\,u_x-A\,u_z+A_z\,u +\textfrac{1}{2}\,A_{zz}\,x^2+A_{tz}\,x\,y,+\textfrac{1}{2}\,A_{tt}\,y^2,
\\
W_1(A) &=& -A\,u_x+A_z\,x+A_t\,y,
\\
W_2(A) &=& A,
\\
X &=& -x\,u_x-y\,u_y + 2\,u,
\\
Y_0 &=& -u_t,
\\
Y_1 &=& - t\,u_t+\textfrac{1}{2}\,(x\,u_x-y\,u_y)-u,
\\
Y_2 &=& -\textfrac{1}{2}\,(t^2\,u_t+t\,x\,u_x-t\,y\,u_y-x\,y)-t\,u,
\\
Z_0 &=& - u_y,
\\
Z_1 &=& -t\,u_y-x,
\end{eqnarray*}
where $A = A(t,z)$ are arbitrary functions. The commutator table of $\mathrm{Sym}_0(\EuScript{E}_2)$ is given
by equa\-ti\-ons
\[[W_i(A), W_j(B)] =
\left\{
\begin{array}{lll}
W_{i+j}(A\,B_z-B\,A_z), &~~~& i+j \le 2
\\
0, && i+j>2,
\end{array}
\right.
\]
\[
[X, W_k(A)] = -k\,W_k(A),
\]
\[
[Y_0, W_k(A)] = W_k(A_t),
\]
\[
[Y_1, W_k(A)] = W_k(t\,A_t+\textfrac{1}{2}\, k\,A),
\]
\[
[Y_2, W_k(A)] = \textfrac{1}{2}\,W_k(t^2\,A_t+k\,t\,A),
\]
\[
[Z_0, W_k(A)] =
\left\{
\begin{array}{lll}
W_{k+1}(A_t), &~~~& k \le 1
\\
0, && k=2
\end{array}
\right.
\]
\[
[Z_1, W_k(A)] =
\left\{
\begin{array}{lll}
W_{k+1}(t\,A_t+k\,A), &~~~& k \le 1
\\
0, && k=2
\end{array}
\right.
\]
\[
\begin{array}{lclcl}
[X, Y_m]  = 0,       & ~~~ &  [X, Z_m] = - Z_m,  & ~~~ & [Z_0, Z_1]=0,
\\   {}
[Y_0, Y_1] = Y_0,  &         &  [Y_0, Y_2]= Y_1,   &         & [Y_1, Y_2] = Y_2,
\\ {}
[Y_0, Z_0] =0, && [Y_1, Z_0] = -\frac{1}{2}\,Z_0, && [Y_2, Z_0] = -\frac{1}{2}\,Z_1,
\\ {}
[Y_0, Z_1] =Z_0, && [Y_1, Z_1] = \frac{1}{2}\,Z_1, && [Y_2, Z_1] =0.
\end{array}
\]
From this table it follows that $\mathrm{Sym}_0(\EuScript{E}_2)$ is the semi-direct product
$\mathfrak{q}_3 =   \mathfrak{q}_{3,\infty} \rtimes \mathfrak{q}_{\diamond}$ of the finite-dimensional Lie
algebra $\mathfrak{q}_{\diamond}$ generated by $X$, $Y_i$, $Z_j$ and the infinite-dimensional ideal
$\mathfrak{q}_{3,\infty}$ generated by $W_0(A)$, $W_1(A)$, $W_2(A)$. We  have
$\mathfrak{q}_{\diamond} = \mathfrak{a} \ltimes (\mathfrak{sl}_2(\mathbb{R}) \ltimes \mathfrak{v})$,
where $\mathfrak{a} =\langle X\rangle$ is one-dimensional Lie algebra,
$\mathfrak{sl}_2(\mathbb{R})  = \langle Y_0, Y_1, Y_2\rangle$, and
$\mathfrak{v} = \langle Z_0, Z_1\rangle$ is two-dimensional  Abelian Lie algebra,
while $\mathfrak{q}_{3,\infty}$ is isomorphic to the  tensor product
$\mathbb{R}_2[h_0] \otimes  \mathfrak{w}[t,z]$, where we denote
$\mathfrak{w}[t,z] = \langle \, t^i z^j \partial_z \,\,\vert\,\, i, j \in \mathbb{N}_{0} \,\rangle$.


\subsection{Maurer--Cartan forms and  the second exotic cohomology group}
\label{MCfs_subsection}

Consider the Maurer--Cartan forms  $\alpha$, $\beta_i$, $i \in \{0, 1, 2\}$, $\gamma_l$, $l \in \{0, 1\}$,
$\theta_{k,i,j}$, $k\in \{0, 1, 2\}$, $i, j \in \mathbb{N}_{0}$, of the Lie algebra $\mathfrak{q}_3$ that are
dual to  the basis $X$, $Y_i$, $Z_l$, $\displaystyle{\frac{1}{i!}\frac{1}{j!}\,W_k(t^iz^j)}$, in other words,
take 1-forms such that there hold
$\alpha(X) =1$,
$\beta_{i}(Y_{i^{\prime}}) = \delta_{ii^{\prime}}$,
$\gamma_{l}(Z_{l^{\prime}}) = \delta_{ll^{\prime}}$,
$\theta_{k,i,j}(W_{k^{\prime}}(t^{i^{\prime}}z^{j^{\prime}}))
= i!\,j!\,\delta_{kk^{\prime}} \,\delta_{ii^{\prime}} \,\delta_{jj^{\prime}}$,
while all the other values of these 1-forms on the elements of the basis are equal to zero. Denote
\[
B = \beta_0 + h_1\,\beta_1+\textfrac{1}{2}\,h_1^2 \beta_2,
\qquad
\Gamma = \gamma_0 + h_1\,\gamma_1,
\]
and consider the formal series of 1-forms
\begin{equation}
\Theta =\sum \limits_{k=0}^{2}
\sum \limits_{i=0}^{\infty}\sum \limits_{j=0}^{\infty} \frac{h_0^k h_1^i h_2^j}{i! j!}\, \theta_{k,i,j},
\label{Theta_FKh4}
\end{equation}
Then the commutator table for the generators of $\mathfrak{q}_3$ yields  the structure equations
\begin{equation}
\fl
\left\{
\begin{array}{lcl}
d\alpha &=& 0,
\\
dB &=& \nabla_1(B) \wedge B,
\\
d\Gamma &=& \alpha \wedge \Gamma + \nabla_1(\Gamma) \wedge B +\textfrac{1}{2}\,\nabla_1(B) \wedge \Gamma,
\\
d\Theta &=& \nabla_2(\Theta) \wedge \Theta
+h_0\,\nabla_0(\Theta) \wedge \left( \textfrac{1}{2}\,\nabla_1(B)+h_0\,\nabla_1(\Gamma)-\alpha\right)
\\
&&
+ \nabla_1(\Theta) \wedge (B+h_0\,\Gamma).
\end{array}
\right.
\label{FKh4_se}
\end{equation}
This system implies the following statement.

\vskip 7 pt

\noindent
{\sc Proposition 2}.
{\it
$H^1(\mathfrak{q}_3) = \langle\, \alpha \,\rangle$,
\[
H_{c\,\alpha}^2(\mathfrak{q}_{\diamond}) =
\left\{
\begin{array}{lcl}
\langle\, [\gamma_0 \wedge \gamma_1] \,\rangle, &~~~& c=2,
\\
\{[0]\}, && c \neq 2,
\end{array}
\right.
\]
and $H_{2\,\alpha}^2(\mathfrak{q}_{\diamond}) \subseteq H_{2\,\alpha}^2(\mathfrak{q}_3)$.
Equation
\begin{equation}
d\sigma = 2\,\alpha \wedge \sigma + \gamma_0 \wedge \gamma_1
\label{FKh4_sigma_eq}
\end{equation}
with unknown 1-form $\sigma$ is compatible with the structure equations
\eqref{FKh4_se}.
System \eqref{FKh4_se}, \eqref{FKh4_sigma_eq} defines the structure equations for a non-central extension
$\widehat{\mathfrak{q}}_3$  of the Lie algebra $\mathfrak{q}_3$.
}

\hfill $\Box$


\subsection{Lax representation of rqsdYM}

Integration of the structure equations \eqref{FKh4_se}, \eqref{FKh4_sigma_eq} gives consequently
\[\fl
\alpha =\frac{d a_0}{a_0},
\qquad
\beta_0 = a_1^2\,dt,
\qquad
\beta_1 = 2\,\frac{d a_1}{a_1}+a_2\,dt,
\qquad
\beta_2 = \frac{1}{a_1^2}\,\left(d a_2 +\frac{a_2^2}{2}\,dt\right),
\]
\[\fl
\gamma_0 = a_0\,a_1\,\left(dy + s \,dt\right),
\qquad
\gamma_1 = \frac{a_0}{a_1}\,\left(ds + \textfrac{1}{2}\,a_2 \,(dy + s\,dt)\right),
\]
\begin{equation}
\fl
\sigma = a_0^2\,\left(dv-s\,dy-\textfrac{1}{2}\,s^2\,dt\right),
\qquad
\theta_{0}= b \,(dz+q_0\,dt),
\label{t000}
\end{equation}
\[\fl
\theta_{1} = \frac{a_0 b}{a_1}\,\left(dx_1 +p_1\,dz + q_0\,dy+q_1\,dt\right)
\]
\[
\fl
\theta_{2} = \frac{a_0^2\,b}{a_1^2}\,
\left(dx_2 + (p_1-s)\,dx_1+p_0\,dz+(q_1-s\,q_0)\,dy+q_2\,dt\right),
\]
where
$a_0 \neq0$,
$a_1 \neq 0$, and $b \neq 0$.
We put    $x_2=u$, $p_1 = -u_{x}+s$, $p_2 = -u_{z}$, $q_1 = -u_{y}+s\,q_0$, $q_2 = -u_{t}$. This yields
\[
\theta_{2}= \frac{a_0^2b}{a_1^2}\,
\left(du -u_{t}dt- u_{x}dx-u_{y}dy-u_{z}dz\right).
\]
Then we alter notation by substituting  
$b = v_{x}v_s^{-1}$, $a_1 =a_0^{-1}v_s^{-1}$,
$q_0 = v_{x}^{-1}\,(v_{y} - \frac{1}{2}\,a_2\,v_s-s)$
in the linear combination
\[
\fl
\mu_1= \sigma -\gamma_1 -\theta_{1}
=
a_0^2\,(du
-a_0^{-1}a_1^{-1}\,(ds+b\,(dx+(s-u_{x})\,dz)
\]
\[
\fl
\qquad\qquad
+(b\,q_0+a_0\,a_1s+\textfrac{1}{2}\,a_2)\,dy
+(b\,(q_0\,s-u_{y})+\textfrac{1}{2}\,s\,(a_2 s+a_0a_1)\,dt)
\]
and obtain
\[
\fl
\mu_1= a_0^2\,(du-v_s\,ds  - v_{x}\,(dx+(s-u_{x})\,dz) -v_{y}\,dy
-(s\,v_{y}-u_{y}v_{x} -\textfrac{1}{2}\,s^2)\,dt).
\]
This is the Wahlquist--Estabrook form for the Lax representation
\begin{equation}
\left\{
\begin{array}{lcl}
v_{z} &=& (s-u_{x})\,v_{x},
\\
v_{t} &=& s\,v_{y} -u_{y}\,v_{x} -\textfrac{1}{2}\,s^2
\end{array}
\right.
\label{FKh4_covering_renamed}
\end{equation}
of equation \eqref{FKh4}.
System \eqref{FKh4_covering_renamed} differs from \eqref{FKh4_covering_lambda} by the change of notation
$s \mapsto \lambda$, $v \mapsto v- \textfrac{1}{2}\,\lambda^2\,t$.


\subsection{Integrable hierarchy associated to rqsdYM}
\label{rdsdYM_hierarchy_subsection}

The Lie algebra  $\mathfrak{q}_3$ admits a series of natural extensions
$\mathfrak{q}_{n+1} = \mathfrak{q}_{n+1,\infty} \rtimes \mathfrak{q}_{\diamond}$ for $n \ge 3$
with  $\mathfrak{q}_{n+1,\infty} = \mathbb{R}_{n+1}[h_0] \otimes \mathfrak{w}[t,z]$.
The  structure equations for the Lie algebra $\mathfrak{q}_{n+1}$ are given by system
\eqref{FKh4_se}, \eqref{FKh4_sigma_eq}, where we put
\begin{equation}
\Theta =\sum \limits_{k=0}^{n}
\sum \limits_{i=0}^{\infty}\sum \limits_{j=0}^{\infty} \frac{h_0^k h_1^i h_2^j}{i! j!}\, \theta_{k,i,j}
\label{Theta_FKh4_n}
\end{equation}
instead of  \eqref{Theta_FKh4} and assume $h_0^k = 0$ for $k > n$. The finite-dimensional part
$\mathfrak{q}_\diamond$ in all the algebras $\mathfrak{q}_{n+1}$ is the same, and  we have
$H^2_{2\alpha_0}(\mathfrak{q}_\diamond) \subseteq H^2_{2\alpha_0}(\mathfrak{q}_{n+1})$ for each $n \ge 3$.
Hence the nontrivial 2-cocycle $\gamma_0 \wedge \gamma_1$ of the differential $d_{2\alpha_0}$ defines a
non-central extension
$\widehat{\mathfrak{q}}_{n+1} = \mathfrak{q}_{n+1,\infty} \rtimes \widehat{\mathfrak{q}}_\diamond$
of the Lie algbera $\mathfrak{q}_{n+1}$. The structure equations for $\widehat{\mathfrak{q}}_{n+1}$ are given
by system \eqref{FKh4_se}, \eqref{FKh4_sigma_eq}  with $\Theta$ defined by \eqref{Theta_FKh4_n}.
To unify notation we rename $t \mapsto y_1$, $x \mapsto t_0$, $y \mapsto y_0$, $z \mapsto t_1$.
Integration of the structure equations for a fixed $n \ge 4$ gives
\[
\theta_{k} =\frac{a_0^k b}{a_1^k}\,
\left(
\sum \limits_{i=0}^{k} \,P_{ki}\,dt_{k-i}
+
\left(\sum \limits_{i=0}^{k-1} \,(-1)^i\,q_{n-1-i}\,s^i\right)  dy_0
+
q_k\,dy_1
\right)
\]
where $k \in \{3, \dots, n\}$, polynomials $P_{ki}$ are defined by \eqref{P_definition}, $t_0$, ... , $t_n$,
$q_0$, ..., $q_n$ are parameters, and forms $\alpha$, $\beta_i$, $\gamma_j$, $\sigma$, $\theta_{0}$, ... ,
$\theta_{2}$ are given by system \eqref{t000}. We put $t_n = u$ and rename
$t_i = x_{n-1-i}$, $i \in \{0, \dots, n-1\}$ to simplify notation in what follows. Then we solve the triangular
linear system of equations $P_{ni} = -u_{x_i}$, $i \in \{0, \dots, n-1\}$, with respect to $p_1$, ... , $p_{n}$
and put $q_n = - u_{y_1}$,
$q_{n-1} =-u_{y_0} -\sum \limits_{i=1}^{n-1} (-1)^i\,q_{n-1-i}\,s^i$.
This yields the contact form
\[
\theta_{n} =\frac{a_0^k b}{a_1^k}\,
\left(du - u_{y_0}  dy_0 -u_{y_1} dy_1 -\sum \limits_{i=0}^{n-1} u_{x_i}\,dx_{i}\right)
\]
Then we consider the linear combination $\mu_n = \sigma -\gamma_1 - \theta_{n-1}$ and put
$b=a_0^{-n-1} v_{x_0} v_s^{1-n}$, $a_1 = a_0^{-1}\,v_s^{-1}$,
$q_{n-2} = (v_{y_0} - s - \frac{1}{2}\,a_2v_s)\,v_{x_0}^{-1}+ \sum \limits_{i=1}^{n-2} (-1)^i\,q_{n-1-i} s^i$.
After this change of notation we obtain
\[
\mu_n =
a_0^2\,\left(
dv
-v_{y_0} dy_0
- (s\,v_{y_0} - u_{y_0}v_{x_0} -\textfrac{1}{2}\,s^2)\, dy_1
-v_s\,ds -v_{x_0}\,dx_0
\phantom{\sum\limits_{i=0}^{n-1}}
\right.
\]
\[
\qquad\qquad
\left.
-
\sum \limits_{i=1}^{n-1} \left(s^i -\sum \limits_{j=0}^{i-1}\,s^{i-j-1}\,u_{x_j}\right) v_{x_0}\,dx_i
\right).
\]
This Wahlquist--Estabrook form defines the Lax representation
\begin{equation}
\left\{
\begin{array}{lcl}
v_{y_1} &=& s\,v_{y_0} -u_{y_0}\,v_{x_0} -\textfrac{1}{2}\,s^2
\\
v_{x_1} &=& (s-u_{x_0})\,v_{x_0},
\\
v_{x_2} &=& (s^2-s\,u_{x_0}-u_{x_1})\,v_{x_0},
\\
&& \dots
\\
v_{x_i} &=&\left(s^i -\sum \limits_{j=0}^{i-1}\,s^{i-j-1}\,u_{x_j}\right)\,v_{x_0},
\\
&& \dots
\\
v_{x_{n-1}} &=&\left(s^{n-1} -s^{n-2} \,u_{x_0} -s^{n-3} \,u_{x_1} - \dots- s\,u_{x_{n-3}} -
u_{x_{n-2}}
\right)\,v_{x_0}.
\end{array}
\right.
\label{FKh4_eq_hierarchy_general_covering}
\end{equation}
Equations for $v_{x_i}$ coincide with system \eqref{Pavlov_eq_hierarchy_general_covering}.
The compatibility conditions of system \eqref{FKh4_eq_hierarchy_general_covering} define the integrable hierarchy
associated to equation \eqref{FKh4}. This hierarchy includes system  $\EuScript{H}_{n-1}$, equation \eqref{FKh4}
written as
\begin{equation}
u_{x_0y_1}= u_{x_1y_0}-u_{y_0}\,u_{x_0x_0}-u_{x_0}\,u_{x_0y_0},
\label{FKh4_renamed}
\end{equation}
and system
\[
u_{x_iy_1} = u_{x_{i+1} y_0} + u_{x_i} u_{x_0 y_0} - u_{y_0}\, u_{x_0 x_i},
\qquad i \in \{0, \dots, n-2\}.
\]


\section{The four-dimensional universal hierarchy equation}

In this section we consider equation $\EuScript{E}_3$ defined by \eqref{4D_UHE}. This equation differs by notation from
\eqref{UHE4_renamed}. We show that the Lax representation \eqref{Pavlov_Stoilov_covering} for equation
\eqref{4D_UHE} can be revealed independently from the hierarchies $\EuScript{H}_n$ in the previous sections
by applying twice the procedure of non-central extension via notrivial exotic 2-cocycles to the symmetry algebra
$\mathrm{Sym}_0(\EuScript{E}_3)$. Furthermore, we find an independent hierarchy associated to $\EuScript{E}_3$.


\subsection{Contact symmetries}

The Lie algebra $\mathrm{Sym}_0(\EuScript{E}_3)$ has generators
\[
\begin{array}{rclcrcl}
W_0(A) &=& -A\,u_y+ A_y\,u-A_t\,z,
&~~&
W_1(A) &=& A,
\\
X_1 &=& -t\,u_t+x\,u_x-u,
&&
X_2 &=& -u_t,
\\
X_3 &=& -2\,x\,u_x-z\,u_z+u,
&&
X_4 &=& -\frac{1}{2}\,z\,u_x-t\,u_z,
\\
X_5 &=& -u_z,
&&
X_6 &=& -u_x,
\end{array}
\]
where $A=A(t,y)$ and $B=B(t,y)$ below are arbitrary functions, and the  commutator table
\[
\{W_i(A), W_j(B)\} = W_{i+j}(A B_y-B A_y),
\]
\[
\begin{array}{rclcrlc}
\{X_1, W_0(A)\} &=& W_0(t A_t),&&
\{X_1, W_1(A)\} &=& W_1(t A_t),
\\
\{X_2, W_0(A)\} &=& W_0(A_t),&&
\{X_2, W_1(A)\} &=& W_1(A_t),
\\
\{X_3, W_0(A)\} &=& 0,&&
\{X_3, W_1(A)\} &=& -W_1(A),
\\
\{X_4, W_0(A)\} &=& -W_1(t A_t),&&
\{X_4, W_1(A)\} &=& 0,
\\
\{X_5, W_0(A)\} &=& -W_1(A_t),&&
\{X_5, W_1(A)\} &=& 0,
\\
\{X_6, W_0(A)\} &=& 0,&&
\{X_6, W_1(A)\} &=& 0,
\end{array}
\]
\[
\begin{array}{rclcrlc}
\{X_1, X_2\} &=& -X_2,&&
\{X_1, X_4\} &=& X_4,
\\
\{X_1, X_6\} &=& X_6,&&
\{X_2, X_4\} &=& X_5,
\\
\{X_3, X_4\} &=& -X_4,&&
\{X_3, X_5\} &=& -X_5,
\\
\{X_3, X_6\} &=& -2\,X_6,&&
\{X_4, X_5\} &=& -\frac{1}{2}\,X_5,
\end{array}
\]
while $\{X_i, X_j\} = 0$ for all the other pairs $i < j$.
The table shows that $\mathrm{Sym}_0(\EuScript{E}_3) = \mathfrak{r}_{2,\infty} \rtimes \mathfrak{r}_\diamond$,
where $\mathfrak{r}_\diamond =  \langle X_m \,\,\vert \,\, m \in \{1, \dots, 6\}  \rangle$
and $\mathfrak{r}_{2,\infty} = \mathbb{R}_1[h_0] \otimes \mathfrak{w}[t,y]$.

Define the Maurer--Cartan forms $\theta_{k,i,j}$ with $k \in \{0, 1\}$, $i,j  \in \mathbb{N}_0$, and
$\beta_m$ with $m \in \{1, \dots, 6\}$, for the Lie algebra $\mathrm{Sym}_0(\EuScript{E}_3)$ as dual 1-forms to
its basis  $\displaystyle{\frac{1}{i!}\,\frac{1}{j!}\,W_k(t^i y^j)}$, $X_m$,
that is, put
$\theta_{k,i,j}(W_{k^\prime}(t^{i^\prime} y^{j^\prime}))
= i!\,j!\,\delta_{k{k^\prime}}\delta_{i{i^\prime}} \delta_{j{j^\prime}}$,
$\theta_{k,i,j}(X_m) = 0$,
$\beta_m(W_k(t^i y^j)) = 0$,
$\beta_m(X_{m^\prime}) = \delta_{m m^\prime}$.
Denote
\begin{equation}
\Theta_k =  \sum \limits_{i=0}^{\infty} \sum \limits_{j=0}^{\infty}
\frac{h_1^i}{i!} \frac{h_2^j}{j!} \theta_{k,i,j},
\label{Theta_UHE4}
\end{equation}
then the system of the Maurer-Cartan structure equations for
$\mathrm{Sym}_0(\EuScript{E}_3)$ is the union of systems
\begin{equation}
\left\{
\begin{array}{lcl}
d\beta_1 &=& 0,
\\
d\beta_2 &=& \beta_1 \wedge \beta_2,
\\
d\beta_3 &=& 0,
\\
d\beta_4 &=& (\beta_3-\beta_1) \wedge \beta_4,
\\
d\beta_5 &=& \beta_3 \wedge \beta_5-\beta_2 \wedge \beta_4,
\\
d\beta_6 &=& (2\,\beta_3- \beta_1) \wedge \beta_6+
\textfrac{1}{2}\,\beta_4 \wedge \beta_5
\end{array}
\right.
\label{4D_UHE_SE_diamond}
\end{equation}
and
\begin{equation}
\left\{
\begin{array}{lcl}
d\Theta_0 &=& \nabla_2(\Theta_0) \wedge \Theta_0 + \nabla_1(\Theta_0) \wedge (\beta_2 +h_1\,\beta_1),
\\
d\Theta_1 &=& \nabla_2(\Theta_1) \wedge \Theta_0+\nabla_2(\Theta_0) \wedge \Theta_1
+(\beta_1-\beta_3) \wedge \Theta_1
\\
&&\quad
+(\beta_2+h_1\,\beta_1) \wedge \nabla_1(\Theta_1)
+(\beta_5+h_1\,\beta_4) \wedge \nabla_1(\Theta_0),
\end{array}
\right.
\label{4D_UHE_SE_infinite}
\end{equation}
where \eqref{4D_UHE_SE_diamond} is the system of the structure equations for
$\mathfrak{r}_\diamond$.


\subsection{Non-central extensions, Maurer--Cartan forms and Lax representation}

Direct computations using the  structure equations \eqref{4D_UHE_SE_diamond}, \eqref{4D_UHE_SE_infinite}
give the following sta\-te\-ment.

\vskip 7pt
\noindent
{\sc Propositon 3}.
{\it
$H^1(\mathrm{Sym}(\EuScript{E}_3)) = \langle \beta_1, \beta_3\rangle$
{\it and}
\[
H^2_{c_1\,\beta_1 + c_2 \,\beta_3}(\mathfrak{r}_\diamond) =
\left\{
\begin{array}{lcll}
\langle
[\beta_2 \wedge \beta_5]
\rangle,
&&
c_1 =1, &c_2 =1,
\\
\langle
[\beta_1 \wedge \beta_2],
[\beta_2 \wedge \beta_3]\rangle,
&~~~~&
c_1 =1,&c_2 =0,
\\
\langle
[\beta_1 \wedge \beta_4],
[\beta_3 \wedge \beta_4]
\rangle,
&&
c_1 =-1,&c_2 =1,
\\
\langle
[\beta_4 \wedge \beta_6]
\rangle,
&&
c_1 =-2,&c_2 =3,
\\
\{[0]\},
&&
\mathrm{otherwise}.&
\end{array}
\right.
\]
Moreover, all the nontrivial exotic 2-cocycles of $\mathfrak{r}_\diamond$ are nontrivial exotic 2-cocycles
of $\mathrm{Sym}_0(\EuScript{E}_3)$ as well. Therefore they define a non-central extension
$\widehat{\mathfrak{r}}_\diamond$ of the Lie algebra $\mathfrak{r}_\diamond$ and hence a non-central extension
$\mathfrak{r}_{2,\infty} \rtimes \widehat{\mathfrak{r}}_\diamond$ of the Lie algebra
$\mathrm{Sym}_0(\EuScript{E}_3)$. The additional Maurer--Cartan forms $\beta_7$, ... , $\beta_{12}$ for the
extended Lie algebra are solutions to system
\begin{equation}
\left\{
\begin{array}{lcl}
d\beta_ 7 &=& (\beta_1+\beta_3) \wedge \beta_7 +\beta_2 \wedge \beta_5,
\\
d\beta_8 &=& \beta_1 \wedge \beta_8 + \beta_1 \wedge \beta_2,
\\
d\beta_9 &=& \beta_1 \wedge \beta_9 + \beta_2 \wedge \beta_3,
\\
d\beta_{10} &=& (\beta_3-\beta_1) \wedge \beta_{10} + \beta_1 \wedge \beta_4,
\\
d\beta_{11} &=& (\beta_3-\beta_1) \wedge \beta_{11} + \beta_3 \wedge \beta_4,
\\
d\beta_{12} &=& (3\,\beta_3-2\,\beta_1) \wedge \beta_{12} + \beta_4 \wedge \beta_6.
\end{array}
\right.
\label{4D_UHE_SE_extended}
\end{equation}
This system is compatible with equations \eqref{4D_UHE_SE_diamond}, \eqref{4D_UHE_SE_infinite}.
}
\hfill $\Box$
\vskip 7 pt

Combining direct integration of the structure equations with Cartan's method of equivalence we  get the
explicit expressions for the Maurer--Cartan forms
\[\fl
\beta_1 = \frac{da_0}{a_0},
\quad
\beta_2 =a_0 \,dt,
\quad
\beta_3 = \frac{da_1}{a_1},
\quad
\beta_4 = \frac{2\,a_1\,ds}{a_0},
\quad
\beta_5 = a_1\,(dz+2\,s\,dt),
\]
\[\fl
\beta_6 = \frac{a_1^2}{a_0}\,(dx + s\,dz + s^2\,dt),
\quad
\beta_{10} = \frac{a_1}{a_0} \,(dw +2\,\ln a_0 \,ds),
\]
\[\fl
\theta_{0} = b\,(dy+r_0\,dt),
\quad
\theta_{1} = \frac{a_1\,b}{a_0}\,(dx_1 + p_1\,dy+r_0\,dz+r_1\,dt).
\]
while the linear combination
\[
\fl
\theta_{1}-\beta_6 =
\frac{a_1b_0}{a_0}\,\left(
dx_1+p_1\,dy-b_0^{-1}\,(a_1\,dx+(a_1s-b_0r_0)\,dz+(a_1s^2-b_0r_1)\,dt)
\right)
\]
must be  a multiple of the contact form $du-u_{t}\,dt-u_{x}\,dx-u_{y}\,dy-u_{z}\,dz$.
Therefore we put $x_1=u$, $p_1 = -u_{y}$, $a_1 = b_0\,u_{x}$, $r_0=-u_{z}+s\,u_{x}$,
$r_1 = -u_{t}+s^2\,u_{x}$.

Our attempts to find a linear combination of the Maurer--Cartan forms $\beta_1$, ... , $\beta_{12}$,
$\theta_{k,i,j}$ have not given a Wahlquist--Estabrook form of any covering over equation \eqref{4D_UHE}.
Therefore we have extended the Lie algebra $\widehat{\mathfrak{r}}_\diamond$ with the structure equations
\eqref{4D_UHE_SE_diamond}, \eqref{4D_UHE_SE_extended} via the same procedure, that is, by finding nontrivial
2-cocycles from $H^2_{c_1 \beta_1 + c_2 \beta_3}(\widehat{\mathfrak{r}}_\diamond)$. Direct computations produce 1
8  such cocycles. The cocycles generate a 18-dimensional non-central extension of the Lie algebra
$\widehat{\mathfrak{r}}_\diamond$. In what follows it is enough to consider one-dimensional extension of
$\widehat{\mathfrak{r}}_\diamond$ generated by the nontrivial 2-cocycle $\beta_{4} \wedge \beta_{10}$ of the
differential $d_{2\beta_3-2\beta_1}$. For the associated Maurer--Cartan form $\beta_{13}$ we have  equation
\begin{equation}
d\beta_{13}=2 \, (\beta_3-\beta_1) \wedge \beta_{13}+\beta_{4} \wedge \beta_{10}.
\label{4D_UHE_beta_13_equation}
\end{equation}
This equation is automatically compatible with system \eqref{4D_UHE_SE_diamond}, \eqref{4D_UHE_SE_infinite},
\eqref{4D_UHE_SE_extended}. Integration of equation \eqref{4D_UHE_beta_13_equation} gives
\[
\beta_{13} = \frac{a_1^2}{a_0^2} \,(dv - 2\,w \,ds).
\]
Consider the linear combination
\[
\fl
\mu_1= \beta_{13}+\beta_5-\beta_6 -\theta_{0}
\]
\[
\fl
=
\frac{b_0^2u_{x}^2}{a_0^2}\,\left(
dv
-2\,w\,ds
-a_0\,dx
-a_0 b_0^{-1}u_{x}^{-2}\,\left(
u_{x}\,(b_0\,s\,u_{x}-a_0)\,dz+a_0\,dy
\right.\right.
\]
\[
\left.\left.
+ (b_0\,s^2\,u_{x}^2-a_0\,(u_{z}+s\,u_{x}))\,dt\right)
\right)
\]
and alter notation by substituting $a_0 = v_{x}$, $b_0 = v_{x}^2u_{x}^{-1}v_{y}^{-1}$, $w=\textfrac{1}{2}\,v_s$.
This gives the Wahlquist--Estabrook form
\[
\fl
\mu_1 = \frac{v_{x}^2}{u_{x}^2v_{y}^2}\,
(dv-v_s\,ds-v_{y}dy-v_{x}dx-(s\,v_{x} - u_{x}\,v_{y})\,dz
\]
\[
-(s^2\,v_{x} - (u_{z}+s\,u_{y})\,v_{y})\,dt)
\]
for the Lax representation
\begin{equation}
\left\{
\begin{array}{lcl}
v_{z} &=& s\,v_{x} - u_{x}\,v_{y},
\\
v_{t} &=& s^2\,v_{x} - (u_{z}+s\,u_{y})\,v_{y}
\end{array}
\right.
\label{Pavlov_Stoilov_covering_renamed}
\end{equation}
of equation \eqref{4D_UHE}. System \eqref{Pavlov_Stoilov_covering_renamed} differs from the Lax representation
\eqref{Pavlov_Stoilov_covering}  by notation.


\subsection{Integrable hierarchy associated to 4D UHE}

To find an integrable hierarchy associated to equation \eqref{4D_UHE} we use the technique of subsections
\ref{Dunajski_hierarchy_subsection} and \ref{rdsdYM_hierarchy_subsection}. Instead of the formal series
\eqref{Theta_UHE4} with $k \in \{0, 1\}$ for fixed $n \ge 2$ consider the formal series
\begin{equation}
\Theta = \sum \limits_{k=0}^n \sum \limits_{i=0}^{\infty} \sum \limits_{j=0}^{\infty}
h_0^k\,\frac{h_1^i}{i!} \,\frac{h_2^j}{j!} \,\theta_{k,i,j}.
\label{Theta_UHE4_extended}
\end{equation}
Then the crucial question is how to generalize system \eqref{4D_UHE_SE_infinite} for the series
\eqref{Theta_UHE4_extended} (note, i.e., that system \eqref{4D_UHE_SE_infinite} does not
involve the Maurer--Cartan form $\beta_6$). We propose to consider system
\begin{equation}
\fl
\begin{array}{lcl}
d\Theta &=&
\nabla_2(\Theta) \wedge \Theta
+ \nabla_1(\Theta) \wedge \left(\beta_2+h_1\,\beta_1+h_0\,(\beta_5+h_1\,\beta_4) +h_0^2\,\beta_6\right)
\\
&&
+h_0\,\nabla_0(\Theta) \wedge \left(\beta_1-\beta_3+\textfrac{1}{2}\,\beta_4\right)
-\textfrac{1}{2}\,\Theta \wedge \beta_4
\end{array}
\label{4D_UHE_SE_infinite_extended}
\end{equation}
that includes \eqref{4D_UHE_SE_infinite} as a subsystem.
To simplify notation in what follows we rename independent variables as  $t=y_2$, $x=y_0$, $y=t_0$, and $z=y_1$.
Integrating equations from \eqref{4D_UHE_SE_infinite_extended} we get
\[
\theta_{k}=  \frac{a_1^k\,b}{a_0^k}\,
\left(
\sum \limits_{i=0}^{k} P_{ki} dt_{k-i}+r_n dy_2
+
\left(\sum \limits_{j=0}^{k-1} (-1)^j r_{k-1-j}s^j\right)\,dy_1
\right.
\]
\[
\qquad\qquad\qquad
\left.
+
\left(\sum \limits_{j=1}^{k-1} (-1)^j \,j\,r_{k-1-j}\,s^{j-1}\right)\,dy_0
\right)
\]
for $k \in \{0, \dots, n\}$, where $r_0$, ... , $r_n$ are parameters,  and polynomials $P_{ki}$ are defined by
\eqref{P_definition}. We take the linear combination $\theta_{n}- \beta_6$, put  $t_n =u$,  solve the
triangular linear system  $P_{ni} = - u_{t_{n-i}}$, $i \in \{0, \dots, n-1\}$, with respect to unknowns
$p_1$, $p_2$, ... , $p_n$, then consequently put
\[
r_n = -u_{y_2}+a_0^{n-1}a_1^{2-n}b_0^{-1}\,s^2,
\]
\[
r_{n-1} = -u_{y_1} - \sum \limits_{j=1}^{n-1} (-1)^j \,r_{n-1-j}\,s^j+a_0^{n-1}a_1^{2-n}b_0^{-1}\,s,
\]
\[
r_{n-2} =
\left\{
\begin{array}{lcl}
\displaystyle{
-u_{y_0} - \sum \limits_{j=2}^{n-1} (-1)^j\,j \,r_{n-1-j}\,s^{j-1}+a_0^{n-1}a_1^{2-n}b_0^{-1},
}&& n\ge 3,
\\
-u_{y_0} +a_0^{n-1}a_1^{2-n}b_0^{-1},
&& n=2,
\end{array}
\right.
\]
and finally rename $t_i = x_{n-1-i}$, $i \in \{0, \dots, n-1\}$. This gives the contact form
\[
\theta_{n} -\beta_6= \frac{a_1^n}{a_0^n}\,\left(
du
- \sum \limits_{i=0}^{n-1} u_{x_i} dx_i
- \sum \limits_{j=0}^{2} u_{y_j} dy_j
\right).
\]
Now we take the linear combination  $\mu_n = \beta_{13}+\beta_5-\beta_6- \theta_{n-1}$, rename
$b= a_0^{n-3}a_1^{3-n}\, v_{x_0}$, and put
\[
a_0 =
\left\{
\begin{array}{lcl}
\displaystyle{
 v_{y_0} - v_{x_0}\,\sum \limits_{j=0}^{n-3} (-1)^j\,(j+1) \,r_{n-3-j}\,s^{j},
}&& n \ge 3,
\\
v_{y_0}, && n = 2.
\end{array}
\right.
\]
Thus we obtain
\[
\fl
\mu_n =
\frac{a_1^2}{a_0^2}\,\left(
dv -v_s\,ds
- v_{y_0} \,(dy_0+s\,dy_1+s^2\,dy_2)
- v_{x_0}\,(u_{y_0} dy_1+(u_{y_1}+s\,u_{y_0})\,dy_2)
\phantom{\sum\limits_{i=0}^{n-1}}
\right.
\]
\[
\left.
-
v_{x_0}\,dx_0-
\sum \limits_{i=0}^{n-1} \left(s^i -\sum \limits_{j=0}^{i-1}\,s^{i-j-1}\,u_{x_j}\right)\,v_{x_0}\, dx_i
\right).
\]
This is the Wahlquist--Estabrook form for the Lax representation
\begin{equation}
\left\{
\begin{array}{lcl}
v_{y_1} &=& s\,v_{y_0} -u_{y_0}\,v_{x_0},
\\
v_{y_2} &=& s^2\,v_{y_0} - (u_{y_1}+s\,u_{y_0})\,v_{x_0},
\\
v_{x_1} &=& (s-u_{x_0})\,v_{x_0},
\\
v_{x_2} &=& (s^2-s\,u_{x_0}-u_{x_1})\,v_{x_0},
\\
&& \dots
\\
v_{x_i} &=&\left(s^i -\sum \limits_{j=0}^{i-1}\,s^{i-j-1}\,u_{x_j}\right)\,v_{x_0},
\\
&& \dots
\\
v_{x_{n-1}} &=&\left(s^{n-1} -s^{n-2} \,u_{x_0} -s^{n-3} \,u_{x_1} - \dots- s\,u_{x_{n-3}} -
u_{x_{n-2}}
\right)\,v_{x_0}.
\end{array}
\right.
\label{4D_UHE_hierarchy_general_covering}
\end{equation}
Equations for $v_{x_i}$ coincide with system \eqref{Pavlov_eq_hierarchy_general_covering}.
The compatibility conditions of system \eqref{4D_UHE_hierarchy_general_covering} define the integrable
hierarchy associated to equation \eqref{4D_UHE}. This hierarchy includes system $\EuScript{H}_{n-1}$,
equation \eqref{4D_UHE} written
as \eqref{UHE4_renamed}, and system
\[
\fl
u_{x_k y_0} =u_{x_{k-2}y_2} +  u_{y_1}\, u_{x_0 x_{k-2}} - u_{x_{k-2}}\, u_{x_0 y_1}
+u_{y_0}\,u_{x_0x_{k-1}}-u_{x_{k-1}}\,u_{x_0y_0},
\]
\[
\fl
u_{x_m y_1} = u_{x_{m-1} y_2} + u_{y_1}\,u_{x_0 x_{m-1}} - u_{x_{m-1}}\, u_{x_0 y_1},
\]
where
$k \in \{2, \dots, n-1\}$,
$m \in \{1, \dots, n-1\}$,
and
$u_{x_0x_i}$ are replaced by the right-hard sides of equations from $\EuScript{H}_{n-1}$. System
\eqref{4D_UHE_hierarchy_general_covering} can be included in the construction of \cite{BogdanovPavlov2017}.


\section{The four-dimensional Mart{\'{\i}}nez Alonso--Shabat equation}

Equation \eqref{MASh4} does not belong to the hierarchies from the previous sections. Its Lax representation
\eqref{MASh4_covering_lambda} was found in \cite{Morozov2014} via the method of \cite{Morozov2009b}.  In this
section we show that \eqref{MASh4_covering_lambda} can be constructed by the technique of the present paper as
well. Also we find the integrable hierarchy generated by the natural extensions of the symmetry algebra
$\mathrm{Sym}_0(\EuScript{E}_4)$ of equation  \eqref{MASh4}.


\subsection{Contact symmetries}

The Lie algebra  $\mathrm{Sym}_0(\EuScript{E}_4)$ is generated by symmetries
\[
V_0(A) = -A \,u_x+ A_x\,u- A_t\,z,
\qquad
V_1(A) = A,
\qquad
W(B) = -B\,u_y,
\]
that depend on arbitrary functions $A = A(t,x)$, $B = B(y,z)$,  and symmetries
\[
X_1 = - u_t,
\qquad
X_2 = - t\,u_t,
\qquad
X_3 = -u_z,
\qquad
X_4 = - z\,u_z.
\]
The commutator table
\[
\fl
\begin{array}{lcl}
\{V_i(A), V_j(\tilde{A})\} = V_{i+j} (A\,\tilde{A}_x - \tilde{A}\,A_x),
&~~~~~&
\{V_i(A), W(B)\} = 0,
\\
\{W(B), W(\tilde{B})\} = W (B\,\tilde{B}_y - \tilde{B}\,B_y),
&&
\{X_1, V_0(A)\} = V_0 (A_t),
\\
\{X_2, V_0(A)\} = V_0 (t\,A_t),
&&
\{X_3, V_0(A)\} = -V_1 (A_t),
\\
\{X_4, V_0(A)\} = 0,
&&
\{X_1, V_1(A)\} = V_1 (A_t),
\\
\{X_2, V_1(A)\} = V_1 (t\,A_t+A),
&&
\{X_3, V_1(A)\} = 0,
\\
\{X_4, V_1(A)\} = -V_1 (A),
&&
\{X_1, W(B)\} = 0,
\\
\{X_2, W(B)\} = 0,
&&
\{X_3, W(B)\} = W(B_z),
\\
\{X_4, W(B)\} = W(z\,B_z),
&&
\{X_1, X_2\} = X_1,
\\
\{X_1, X_3\} = 0,
&&
\{X_1, X_4\} = 0,
\\
\{X_2, X_3\} = 0,
&&
\{X_2, X_4\} = 0,
\\
\{X_3, X_4\} = X_3.
&&
\end{array}
\]
implies that
$\mathrm{Sym}_0(\EuScript{E}_4) =
\mathfrak{s}_{2,1}= \mathfrak{s}_{2,1,\infty} \rtimes \mathfrak{s}_\diamond$,
where
$\mathfrak{s}_{2,1,\infty}$ $=$
$\left(\mathbb{R}_2[h_0]\otimes \mathfrak{w}[t,x]\right) \oplus \mathfrak{w}[z,y]$ and
$\mathfrak{s}_\diamond$  $=$  $\langle X_1, X_2\rangle \oplus \langle X_3, X_4\rangle$
(direct sums of Lie algebras).


\subsection{Maurer--Cartan forms and the second exotic cohomology group}
Let the Maurer--Cartan forms
$\theta_{k,i,j}$, $\omega_{ij}$, $\alpha_0$, $\alpha_1$, $\beta_0$, $\beta_1$
be defined by requirement that there hold
$\theta_{k,i,j}(V_{k^\prime}(t^{i^\prime}x^{j^\prime}))
=i!\,j!\,\delta_{k k^\prime}\delta_{i i^\prime}\delta_{j j^\prime}$,
$\omega_{i,j}(W(y^{i^\prime}z^{j^\prime}))=\delta_{i i^\prime}\delta_{j j^\prime}$,
$\alpha_{m}(X_{m^\prime+1})=\delta_{m m^\prime}$,
$\beta_{m}(X_{m^\prime+3})=\delta_{m m^\prime}$,
while all the other values of these forms on the generators of $\mathrm{Sym}_0(\EuScript{E}_4)$ are equal
to zero. Then the structure equations for $\mathrm{Sym}_0(\EuScript{E}_4)$ are given by systems
\begin{equation}
\fl
\left\{
\begin{array}{lcl}
d\alpha_0 &=& 0,
\\
d\alpha_1 &=& \alpha_0 \wedge \alpha_1,
\\
d\beta_0 &=& 0,
\\
d\beta_1 &=& \beta_0 \wedge \beta_1
\end{array}
\right.
\label{MASh4_SE_diamond}
\end{equation}
and
\begin{equation}
\fl
\left\{
\begin{array}{lcl}
d\Theta &=& \nabla_2(\Theta) \wedge \Theta+ \nabla_1(\Theta) \wedge (\alpha_1+h_1\,\alpha_0-h_0\,\beta_1)
\\
&&
+ h_0\,\nabla_0(\Theta) \wedge (\alpha_0 - \beta_0),
\\
d\Omega &=& \nabla_3(\Omega) \wedge \Omega +\nabla_4(\Omega) \wedge (\beta_1 + h_4\,\beta_0).
\end{array}
\right.
\label{MASh4_SE_infinite}
\end{equation}
where
\begin{equation}
\Theta = \sum \limits_{k=0}^{1} \sum \limits_{i=0}^{\infty} \sum \limits_{j=0}^{\infty}
h_0^k\,\frac{h_1^i}{i!}\,
\frac{h_2^j}{j!}\,
\theta_{k,i,j},
\qquad
\Omega = \sum \limits_{i=0}^{\infty} \sum \limits_{j=0}^{\infty}
\frac{h_3^i \,h_4^j}{i !\, j!}\, \omega_{i,j}.
\label{MASh4_Theta_1}
\end{equation}
System \eqref{MASh4_SE_diamond} defines the structure equations for the Lie algebra $\mathfrak{s}_\diamond$.
Direct computations show that the following statement holds.
\vskip 7 pt
\noindent
{\sc Proposition 4.}
{\it
$H^1(\mathrm{Sym}_0(\EuScript{E}_4)) = \langle \alpha_0, \beta_0 \rangle$,
\[
H^2_{c_1\alpha_0+c_2\beta_0}(\mathfrak{s}_\diamond) =
\left\{
\begin{array}{lcl}
\langle [\alpha_0 \wedge \alpha_1], [\beta_0 \wedge \alpha_1]\rangle,
&~~&  c_1=1, c_2=0,
\\
\langle [\alpha_0 \wedge \beta_1], [\beta_0 \wedge \beta_1]\rangle,
&~~&  c_1=0, c_2=1,
\\
\langle [\alpha_1 \wedge \beta_1]\rangle,
&~~&  c_1=1, c_2=1,
\\
\{[0]\}, && \mathrm{otherwise},
\end{array}
\right.
\]
and
$H^2_{c_1\alpha_0+c_2\beta_0}(\mathfrak{s}_\diamond)
\subseteq H^2_{c_1\alpha_0+c_2\beta_0}(\mathrm{Sym}_0(\EuScript{E}_4))$ for
$(c_1,c_2) = (1,0)$,
$(c_1,c_2) = (0,1)$,
$(c_1,c_2) = (1,1)$.
Equations
\begin{equation}
\left\{
\begin{array}{lcl}
d\gamma_1 &=& \alpha_0 \wedge \gamma_1 + \alpha_0 \wedge \alpha_1,
\\
d\gamma_2 &=& \alpha_0 \wedge \gamma_2 + \beta_0 \wedge \alpha_1,
\\
d\gamma_3 &=& \beta_0 \wedge \gamma_3 + \alpha_0 \wedge \beta_1,
\\
d\gamma_4 &=& \beta_0 \wedge \gamma_4 +\beta_0 \wedge \beta_1,
\\
d\gamma_5 &=& (\alpha_0 +\beta_0) \wedge \gamma_5 + \alpha_1 \wedge \beta_1.
\end{array}
\right.
\label{MASh4_SE_extension}
\end{equation}
with unknown 1-forms $\gamma_1$, ... , $\gamma_5$
are compatible with the structure equations
\eqref{MASh4_SE_diamond}, \eqref{MASh4_SE_infinite}
of $\mathrm{Sym}_0(\EuScript{E}_4)$. System \eqref{MASh4_SE_diamond}, \eqref{MASh4_SE_infinite},
\eqref{MASh4_SE_extension} defines five-dimensional non-central extension for this Lie algebra.
}
\hfill $\Box$


\subsection{Lax representation of the 4D MASh equation}

Integration of the structure equations \eqref{MASh4_SE_diamond}, \eqref{MASh4_SE_infinite},
\eqref{MASh4_SE_extension} gives the Maurer--Cartan forms
$\alpha_0 = db_1/b_1$, $\alpha_1 = b_1\,dt$, $\beta_0 = db_2/b_2$,
$\beta_1 = b_2\,dz$,
$\gamma_1 = b_1\,(dv_1+\ln b_1\, dt)$,
$\gamma_2 = b_1\,(dv_2+\ln b_2\, dt)$,
$\gamma_3 = b_2\,(dv_3+\ln b_1\, dz)$,
$\gamma_4 = b_2\,(dv_4+\ln b_2\, dz)$,
$\theta_{0} = a_1\,(dx + q_0\,dt)$,
$\theta_{1} = a_1\,b_2\,(dx_1+p_1 \,dx -q_0\,dz +q_1\,dt)$,
$\omega=\omega_{0,0} = a_2\,(dy+s_0\,dt)$,
where $b_1 > 0$, $b_2 >0$, $a_1 \neq 0$, $a_2\neq 0$. The results of Cartan's method of equivalence
show that the linear combination $\theta_{1}-\omega$ is a multiple of the contact form
$du-u_{t}dt-u_{x}dx-u_{y}dy-u_{z}dz$. Therefore we put
$x_1=u$, $p_1 = -u_{x}$, $q_1 = -u_{t}$, $a_2 = a_1\,b_2\,b_1^{-1}\,u_{y}$,
$q_0 = u_{z} - s_0\,u_{y}$. Then we consider the linear combination
\[
\fl
\mu_1= \gamma_4 -\gamma_3 +\alpha_1-\beta_1-\theta_{0}-\omega
\]
\[\fl
=
b_2\,(dv_4-dv_3-a_1b_2^{-1}\,dx -a_1b_2^{-1}\,(u_z-s_0\,u_y)\,dt
\]
\[
\fl\qquad\qquad
-a_1b_1^{-1}\,u_y\,dy
-(a_1b_1^{-1}s_0\,u_y+\ln b_1 - \ln b_2)\,dz
+b_1^{-1}b_2^{-1}\,db_1-b_2^{-2}\,db_2).
\]
We substitute
$v_4 = v+v_3$,
$b_2 = s\,b_1$,
$a_1 = b_1\,s\,v_{x}$,
$s_0=u_y^{-1}\,(u_z-v_t\,v_x^{-1})$,
and then put $b_1=s^{-2}\,v_s^{-1}$. This yields the Wahlquist--Estabrook form
\[
\fl
\mu_1=
\frac{1}{s\,v_s}\,(dv-v_s\,ds -v_{t}\,dt -v_{x}\,dx - s\,u_{y}\,v_{x}\,dy
- (s\,(u_{z}v_{x} - v_{t})-\ln s)\,dz).
\]
of the Lax representation
\begin{equation}
\left\{
\begin{array}{lcl}
v_{y} &=& s\,u_{y}\,v_{x},
\\
v_{z} &=& s(u_{z} \,v_{x} - v_{t}) - \ln s
\end{array}
\right.
\label{MASh4_covering_renamed}
\end{equation}
for  equation \eqref{MASh4}.
The change of variables  $s\mapsto \lambda$, $v \mapsto v-z\,\ln \lambda$,
transforms system \eqref{MASh4_covering_renamed} to the form
\eqref{MASh4_covering_lambda}.


\subsection{Integrable hierarchy associated to 4D MASh equation}
\label{MASh_hierarchy_subsection}

Consider a series of natural extensions
\[
\mathfrak{s}_{n,m} =
\left(\mathbb{R}_n[h_0]\otimes \mathfrak{w}[t,x]\right) \oplus
\left(\mathbb{R}_m[h_0]\otimes \mathfrak{w}[z,y]\right)
\rtimes \mathfrak{s}_\diamond
\]
of the Lie algebra $\mathrm{Sym}_0(\EuScript{E}_4)=\mathfrak{s}_{2,1}$.
We replace the series for $\Theta$ from \eqref{MASh4_Theta_1} in \eqref{MASh4_SE_infinite} by
\begin{equation}
\Theta = \sum \limits_{k=0}^{n} \sum \limits_{i=0}^{\infty} \sum \limits_{j=0}^{\infty}
h_0^k\,\frac{h_1^i}{i!}\,
\frac{h_2^j}{j!}\,
\theta_{k,i,j}.
\label{MASh4_Theta_n}
\end{equation}
Then the key question there is how to generalize equations \eqref{MASh4_SE_infinite} for the series
\begin{equation}
\Omega = \sum \limits_{k=0}^{m} \sum \limits_{i=0}^{\infty} \sum \limits_{j=0}^{\infty}
h_0^k\,\frac{h_3^i}{i!}\,
\frac{h_4^j}{j!}\,
\omega_{k,i,j}
\label{MASh4_Omega_m}
\end{equation}
instead of the series for $\Omega$ from \eqref{MASh4_Theta_1}.

We propose to define the Lie algebra $\mathfrak{s}_{n,m}$ by the structure equations that include system
\eqref{MASh4_SE_diamond} and system
\begin{equation}
\fl
\left\{
\begin{array}{lcl}
d\Theta &=& \nabla_2(\Theta) \wedge \Theta+ \nabla_1(\Theta) \wedge (\alpha_1+h_1\,\alpha_0-h_0\,\beta_1)
\\
&&
+ h_0\,\nabla_0(\Theta) \wedge (\alpha_0 - \beta_0),
\\
d\Omega &=& \nabla_3(\Omega) \wedge \Omega +\nabla_4(\Omega) \wedge (\beta_1 + h_4\,\beta_0-h_0\,\alpha_1)
\\
&&
+ h_0\,\nabla_0(\Omega) \wedge (\beta_0 - \alpha_0)
\end{array}
\right.
\label{extended_MASh4_SE}
\end{equation}
with $\Theta$ and $\Omega$ given by \eqref{MASh4_Theta_n} and \eqref{MASh4_Omega_m}. We have
$H^2_{c_1\alpha_0+c_2\beta_0}(\mathfrak{s}_\diamond) \subseteq H^2_{c_1\alpha_0+c_2\beta_0}(\mathfrak{s}_{n,m})$
for $(c_1,c_2) = (1,0)$, $(c_1,c_2) = (0,1)$, $(c_1,c_2) = (1,1)$, therefore system \eqref{MASh4_SE_extension}
defines a non-central extension of the Lie algebra $\mathfrak{s}_{n,m}$.

For simplicity of computations we take $m=n$ in what follows, also we rename
the independent variables as $t = t_0$, $x=x_0$.
Then we find explicit ex\-pres\-si\-ons for the Maurer--Cartan forms $\theta_k=\theta_{k,0,0}$,
$\omega_k = \omega_{k,0,0}$, $k \in \{0, \dots, n\}$ with fixed $n \ge 3$.

Integrating the structure equations \eqref{extended_MASh4_SE}
we obtain
\[
\theta_{k} = a_1 b_1^{-k} b_2^{k}\,\left(dx_k +\sum \limits_{i=1}^k p_i dx_{k-i} - q_{k-1}\,dz+q_n \,dt\right),
\]
\[
\omega_{k} = a_2 b_1^{k} b_2^{-k}\,\left(dy_k +\sum \limits_{i=1}^k r_i dy_{k-i} - s_{k-1}\,dt+q_n \,dz\right),
\]
where $p_i$, $q_i$, $r_i$, $s_i \in \mathbb{R}$ are free parameters.
Then we put consequently
$x_n=u$,
$b_2=b_1\,s$,
$p_1 = - u_{x_{n-1}}+s^{-1}$,
$p_{k} = -u_{x_{n-k}}+2^{k-1}\,s^{-k} - \sum \limits_{j=1}^{k-1} 2^{j-1}\,s^{-j}\,u_{x_{n+j-k}}$,
for $k \in \{2, \dots, n\}$,
$r_1 = u_{y_{n-2}}u_{y_{n-1}}^{-1} - s$,
$r_k = \left(u_{y_{n-k-1}}- s\,u_{y_{n-k}}\right)\,u_{y_{n-1}}^{-1}$ for $k \in \{2, \dots, n-1\}$,
$q_n = - u_t+s^{-1}\,\left(-u_z+2\,q_{n-1}+s_{n-1}\,u_{y_{n-1}}\right)$,
 and
$q_0 = -s^{n-1} \,\left(u_z - u_{y_{n-1}}\,\sum \limits_{j=0}^{n-1} s^{n-1-j}\,s_j\right)
+s^{n-1}\,q_{n-1}-\sum \limits_{k=0}^{n-2} s^k\,q_k$.
This yields the contact form
\[
\fl
\qquad
\theta_{n} -
\sum \limits_{k=0}^{n-1}(\theta_k+\omega_k)
 = a_1\,b_1^{-n}\,b_2^{-n}\,
\left(du-\sum \limits_{k=1}^{n-1} (u_{t_k} dt_i+u_{x_k} dx_k)-u_{y} dy -u_{z} dz\right).
\]
Then we consider the linear combination
\[
\mu_n = \gamma_4-\gamma_3+\alpha_0-\beta_0
-\theta_{n-1} +\sum \limits_{k=0}^{n-3} (n-k-2)\,\theta_k -\sum \limits_{k=0}^{n-1}(n-k)\,\omega_k
\]
and put consequently $v_4 = v+v_3$, $a_1=b_1\,s^{2-n}\,v_{x_{n-1}}$, $b_1=s^{-2}\,v_s$,
\[
q_1 = s^{n-2}\,\left(v_{t}\,v_{x_{n-1}}^{-1}-(n-2)\,u_z+(n-3)\,s_{n-1}\right)
-\sum \limits_{k=2}^{n-2}\,k\,s^{k-1}\,q_k
\]
\[\qquad\qquad
+u_{y_{n-1}}\,\sum \limits_{k=0}^{n-1} (2\,n-k-3)\,s^{2\,n-k-3}\,s_k
\]
in the case $n>3$ and
\[
q_1 = s\,\left(v_{t}\,v_{x_{2}}^{-1}-\,u_z\right)
+u_{y_{2}}\,\sum \limits_{k=0}^{2} (3-k)\,s^{3-k}\,s_k
\]
when $n=3$.
Finally we rename $x_{k} \mapsto x_{n-k-1}$, $y_{k} \mapsto y_{n-k-1}$ for $k \in \{0, \dots , n-1\}$.
This gives
the Wahlquist--Estabrook form
\[
\fl
\mu_n =
\frac{1}{s\,v_s}\,\left(
dv-v_s\,ds -v_t\,dt-v_{x_0} \,dx_0- \left(s\,(u_z\,v_{x_0}-v_t) - \ln s\right)\,dz
\phantom{\sum\limits_{i=0}^{n-1}}
\right.
\]
\[
\left.
- \sum \limits_{k=0}^{n-1}\left(\sum \limits_{j=0}^k s^{k-j}\,u_{y_j}\right) \, v_{x_0}\, dy_k
-\sum \limits_{m=1}^{n-1} \left(s^{-m} -\sum \limits_{j=0}^{m-1}\,s^{1-m-j}\,u_{x_j}\right)\, v_{x_0}\,dx_m
\right)
\]
for the Lax representation
\begin{equation}
\left\{
\begin{array}{lcl}
v_{x_1} &=& \left(s^{-1} -u_{x_0}\right)\, v_{x_0},
\\
&& \dots
\\
v_{x_m} &=& \left(s^{-m} -\sum \limits_{j=0}^{m-1}\,s^{1-m-j}\,u_{x_j}\right)\, v_{x_0},
\\
&& \dots
\\
v_{x_{n-1}} &=& \left(s^{1-n} -s^{2-n}\,u_{x_0} -s^{3-n} \,u_{x_1} - \dots -u_{x_{n-2}}\right)\, v_{x_0},
\\
v_{y_0} &=& s\,u_{y_0} \, v_{x_0},
\\
&& \dots
\\
v_{y_k} &=& \left(\sum \limits_{j=0}^k s^{k+1-j}\,u_{y_j}\right) \, v_{x_0},
\\
&& \dots
\\
v_{y_{n-1}} &=& \left(\sum \limits_{j=0}^k u_{y_{n-1}}+s\,u_{y_{n-2}}+\dots+ s^{n-2}\,u_{y_0}\right) \, v_{x_0},
\\
v_z &=& s\,(u_z\,v_{x_0}-v_t) - \ln s.
\end{array}
\right.
\label{MASh4_eq_hierarchy_general_covering}
\end{equation}
Equations for $v_{x_m}$ differ from system \eqref{Pavlov_eq_hierarchy_general_covering} by the change
$s\mapsto s^{-1}$. The compatibility conditions of system
\eqref{MASh4_eq_hierarchy_general_covering}
define the integrable hierarchy associated to equation
\eqref{MASh4}.
This hierarchy includes system $\EuScript{H}_{n-1}$ as well as  equation \eqref{MASh4} written in the form
\[
u_{ty_0} = u_z\,u_{x_0y_0} -u_{y_0}\,u_{x_0z},
\]
systems
\begin{equation}
u_{tx_k} = -u_{x_{k+1}z} +u_z\,u_{x_0x_k} -u_{x_k}\,u_{x_0z},
\label{utx_k_series}
\end{equation}
\begin{equation}
u_{ty_m} = -u_{y_{m-1}z} +u_z\,u_{x_0y_m} -u_{y_m}\,u_{x_0z},
\label{utym_series}
\end{equation}
\begin{equation}
u_{x_{k+1}y_0} =u_{y_0}\,u_{x_0x_k}-u_{x_k}\,u_{x_0y_0},
\label{uxky0_series}
\end{equation}
\begin{equation}
u_{x_{k+1}y_m} =u_{x_ky_{m-1}} + u_{y_m}\,u_{x_0x_k}-u_{x_0}\,u_{x_ky_m}
\label{uxk1ym_series}
\end{equation}
\begin{equation}
u_{y_0y_k}= u_{y_0}\,u_{x_0 y_{k+1}}-u_{y_{k+1}}\,u_{x_0 y_{0}}
\label{uy0yk_series}
\end{equation}
with $k \in \{0, \dots, n-2\}$, $m \in \{1, \dots, n-1\}$,
and system
\begin{equation}
u_{y_iy_j}= u_{y_{i-1}y_{j+1}}+u_{y_i}\,u_{x_0 y_{j+1}}-u_{y_{j+1}}\,u_{x_0 y_{i}}
\label{uyiyj_series}
\end{equation}
with $ i \in \{1, \dots , n-2\}$, $j \in \{i, \dots , n-2\}$.

Some equations from systems \eqref{utx_k_series} --- \eqref{uyiyj_series} differ by notation from
equations \eqref{FKh4} --- \eqref{UHE3}.
Equations \eqref{utx_k_series}, \eqref{uxk1ym_series} with $k=0$,
equation \eqref{uyiyj_series} with $i=j=1$,  and
equations \eqref{uxky0_series},  \eqref{uy0yk_series} with $k>0$
correspond to equations \eqref{FKh4}, \eqref{4D_UHE}, and \eqref{MASh4}, respectively.
Equations \eqref{utx_k_series} with $k>0$ or equations \eqref{utym_series},
and equation \eqref{uy0yk_series} with $k=0$ agree with equations
\eqref{MASh5} and \eqref{UHE3}, respectively.
Equation \eqref{uxky0_series} with $k=0$ is the 3-dimensional rdDym equation,
\cite{Blaszak,Pavlov2003,Morozov2009b,BKMV2016},
When the right-hand sides of equations  \eqref{uxk1ym_series}, \eqref{uyiyj_series} inlcude the terms
$u_{x_ky_{m-1}}$, $u_{x_ky_m}$, and $u_{y_{i-1}y_{j+1}}$ from the left-hand sides of equations from the same
systems, these terms have to be replaced using the corresponding equations. This yields equations of
increasing dimensions.


\section{Conclusion}
In the present paper we have shown that the method of \cite{Morozov2017,Morozov2018a} is applicable to
Lax re\-pre\-sen\-ta\-ti\-ons with non-removable parameters, in particular, the Lax representations for
equations \eqref{Pavlov_eq} --- \eqref{MASh4} can be derived from
the non-central extensions of contact symmetry algebras of these equations. In all the  examples the symmetry
al\-ge\-b\-ras have the specific structure of the semi-direct product of an infinite-dimensional ideal and a
non-Abelian finite-di\-men\-si\-o\-nal Lie subalgebra. The cohomological properties of the fi\-ni\-te-dimensional
subalgebras turn out to be sufficient to reveal non-central extensions that define the Lax representations.

For the considered equations the infinite-dimensional ideals of the symmetry al\-geb\-ras are either of the
form of tensor products of the algebra of trun\-ca\-ted polynomials with an
infinite-dimensional Lie algebra, or contains such tensor products as direct summands. The natural procedure of
increasing the degree of the trun\-ca\-ted polynomials produces a series of natural extensions of the symmetry
algebras of the {\sc pde}s under the study. These extensions inherit the nontrivial exotic 2-cocycles and thus
admit non-central extensions generated by these cocycles. We have shown that this procedure gives integrable
hierarchies associated with equations \eqref{Pavlov_eq}, \eqref{FKh4}, \eqref{4D_UHE},
and \eqref{MASh4}.

It is natural to ask whether the method is be applicable in the case when the sym\-met\-ry algebra of
the {\sc pde} has more complicated structure.  Also, we expect that the pro\-po\-sed technique will be
helpful in describing multi-component integrable ge\-ne\-ra\-li\-za\-ti\-ons of integrable {\sc pde}s,
\cite{Dunajski2002,ManakovSantini2006,Bogdanov2010,Morozov2012,KruglikovMorozov2015}.
We intend to address these issues in the further study.


\section*{Acknowledgments}

This work was partially supported by the Faculty of Applied Mathematics of AGH UST statutory tasks within
subsidy of Ministry of Science and Higher Education.

I am very grateful to I.S. Krasil${}^{\prime}$shchik for useful discussions.
I thank L.V. Bog\-da\-nov, M.V. Pavlov,  and P. Zusmanovich for important remarks.


\end{document}